\documentclass[preprint]{aastex63}

\def\bfc{}

\def\bfred{}

\def\cm3{cm$^{-3}$}

\usepackage{graphicx}
\usepackage{epsf}
\usepackage{natbib}

\newcommand{\simlt}{\lower.5ex\hbox{$\; \buildrel < \over \sim \;$}}

\begin{document}

\title{Ionized carbon around IRC +10216}

%\shorttitle{Halo around Evolved Stars}

\author[0000-0001-8362-4094]{William T. Reach}
\affil{Universities Space Research Association, MS 232-11, NASA Ames Research Center, Moffett Field, CA 94035, USA}
\email{wreach@sofia.usra.edu}

\author[0000-0003-0522-5789]{Maxime Ruaud}
\affil{Carl Sagan Center, SETI Institute, Mountain View, CA 94035, USA}
\affil{NASA Ames Research Center, Moffett Field, CA 94035, USA}
\email{mruaud@seti.org}

\author[0000-0002-5135-8657]{Helmut Wiesemeyer}
\affil{Max Planck Institute for Radio Astronomy, Bonn, Germany}
\email{hwiese@mpifr-bonn.mpg.de}

\author[0000-0001-5389-0535]{Denise Riquelme}
\affil{Max Planck Institute for Radio Astronomy, Bonn, Germany}
\email{riquelme@mpifr-bonn.mpg.de}

\author[0000-0002-6488-8227]{Le Ngoc Tram}
\affil{Universities Space Research Association, MS 232-11, NASA Ames Research Center, Moffett Field, CA 94035, USA}
\affil{Max Planck Institute for Radio Astronomy, Bonn, Germany}

\author[0000-0002-3518-2524]{Jose Cernicharo}
\affil{Instituto de Física Fundamental (IFF-CSIC), C/Serrano 121, 28006 Madrid, Spain}
\email{jose.cernicharo@csic.es}

\author[0000-0001-5510-2424]{Nathan Smith}
\affil{Steward Observatory, University of Arizona, 933 N. Cherry Ave., Tucson, AZ 85721, USA}
\email{nathans@as.arizona.edu}

\author[0000-0003-4195-1032]{Edward T. Chambers}
\affil{Universities Space Research Association, MS 232-11, NASA Ames Research Center, Moffett Field, CA 94035, USA}
\email{echambers@sofia.usra.edu}

\begin{abstract}
Asymptotic giant branch (AGB)
stars create a rich inventory of molecules in their envelopes as they lose mass during later stages of their evolution. 
These molecules cannot survive the  conditions in interstellar space, where they are exposed to ultraviolet photons of the interstellar radiation field.
As a result, daughter molecules are the ones injected into space, and a halo of those molecules is predicted to exist around cool  evolved stars.
The most abundant molecule in the envelopes other than H$_2$ is CO, which dissociates into C that is rapidly ionized into C$^+$ 
in a halo around the star that is optically thin to the interstellar radiation field.
We develop the specific predictions of the ionized carbon halo size and column density for the well-studied, nearby star IRC +10216.
We compare those models to  observations of the [C II] 157.7 $\mu$m far-infrared fine-structure line using SOFIA and Herschel.
The combination of bright emission toward the star and upper limits to extended [C II] is
inconsistent with any standard model.
The presence of [\ion{C}{2}] toward the star requires some dissociation and ionization
in the inner part of the outflow, possibly due to a hot companion star.
The lack of extended [\ion{C}{2}] emission requires that daughter
products from CO photodissociation in the outer envelope remain cold. 
The [\ion{C}{2}] profile toward the star is asymmetric, with the blue-shifted absorption
due to the cold outer envelope. 
\end{abstract}

%\keywords{dust, ISM: abundances, ISM: atoms, ISM: clouds, ISM: general,  ISM: molecules}

\section{Introduction}

After their main-sequence lifetimes, low-mass stars lose considerable mass, primarily near the end of their lifetime as giants \citep{wilson00,smith14}.
The mass loss affects the stars' evolution and the composition of the regions around them, with star 
forming regions capable of significant enrichment on timescales less than 100 Myr \citep{mcwilliam97}. 
The material shed by these evolved stars enriches the interstellar medium with nucleosynthetic products that depend on the 
temperature and composition of the portion of the star's atmosphere that is ejected.

IRC+10216 (CW Leonis) is a well-studied asymptotic giant branch (AGB) star that is exceptionally bright \citep{becklin69} owing to
its high luminosity  \citep[1600 $L_\odot$; ][]{menten12} and nearby distance \citep[123 pc; ][]{groenewegen12}.
The star is shedding mass at $2\times 10^{-5}$ $M_\odot$~yr$^{-1}$ {\bfred\citep{fonfria22}}, leading to a rich circumstellar chemistry close to the star \citep{agundez10,agundez12},
followed by an expanding envelope prominent in CO and far-infrared continuum including
shells \citep{fong06,cernicharo15,decin12}. 
Farther from the star, a bow shock has formed where the expanding envelope collides with the surrounding interstellar medium  \citep{sahai10}.

AGB stars are a significant source of carbon and oxygen,
from nucleosynthesis in the stellar interior by the CNO cycle, then dredged up 
during stellar evolution, leading to enrichment of the ISM \citep{busso99}.
The carbon is  
launched into the wind by drag on solid particles
formed just outside the photosphere \citep{fonfria21}. 
Stars with C/O$>1$ 
require dredge-ups that occur only if the initial mass was $>1.5 M_\odot$ \citep{hofner18}.
For stars with initial mass 1.5--2 $M_\odot$, the carbon abundance in the wind is
$1\times 10^{-3} < {\rm [C/H]} < 5\times 10^{-3}$ \citep{karakas10}.
In contrast, the carbon abundance in the ISM, as inferred from abundances in B-star photospheres, is [C/H]=$2.1\times 10^{-4}$ \citep{nieva12}. Thus 1.5--2 $M_\odot$
carbon stars eject
material that is enriched in carbon by 5--25 times compared to the present-day ISM. 
The abundance of carbon-bearing molecules (primarily CO) is only 
[C/H]=$\frac{1}{2}$[C/H$_2$]=$4\times 10^{-4}$ \citep{saberi19}, % BETTER REF?
so the bulk of the carbon, between 60\% and 92\% for the 1.3--2 $M_\odot$ initial mass models,
is not in molecules and is presumably in solid form.

\def\extra{
For comparison the Sun has a C abundance 50\% higher \citep{GS98} than the B-star average, which should be sampling
the ISM over 4 Gyr ago. The higher C abundance in the Sun has been interpreted as potential local
enrichment from forming near massive stars \citep{snow06,slavin06}. 
But the enrichment we infer in the outflow from IRC+10216 is much more than the slightly enhanced solar abundance. 
IRC+10216 is in its AGB stage, so progenitor is of intermediate mass, and it should have formed earlier than present B
stars and would have had a lower initial [C/H] than those stars. 

\begin{figure}
    \centering
    \includegraphics[width=4in]{yieldfig}
    \caption{Abundance of C predicted in the stellar winds of stars as a function of their initial mass \citep{karakas10}, if the initial metallicity of the star is $Z_0$ listed in the inset legend. Dots connected by
    lines are for the unmixed atmosphere models, and isolated dots are for the mixed models.
    Horizontal lines indicate the [C/H] abundance of the present ISM \citep{nieva12}, 
    that used in modeling CO in stellar envelopes \citep{saberi19}, and from the only reasonably successful model from this paper. Most of the C enriching the ISM from carbon stars is not from dissociated CO but rather from a 
    shorter-lived source or elemental C that remains neutral and/or cold in the outflow.
    }
    \label{yieldfig}
\end{figure}
}

In this paper we calculate the size, column density, and brightness of far-infrared fine structure
line emission from the circumstellar envelope extending all the
way to where the CO gas is dissociated. 
The size of the CO envelope is a basic observable property of evolved star mass loss and is an accessible tracer of the mass-loss rate with high-resolution mm-wave imaging \citep{ramstedt20}. 
Our model predicts the properties of the halo of carbon that should occur outside the CO envelope.
We compare the model to far-infrared observations of [\ion{C}{2}] with Herschel and SOFIA.

\section{Model for halo of ionized carbon}

The `standard model' for steady mass loss from a star at rate $\dot{M}$, expanding with constant velocity $v$, 
produces a cloud of material with  density $n$ that decreases as the inverse square of distance $r$ to the star:
\begin{equation}
    n = \frac{\dot{M}}{4 \pi \mu m_{\rm H} r^2 v } \label{massloss}
\end{equation}
where $\mu=2.21$ is the mean mass for neutral, molecular gas (predominantly H$_2$ and He).
For a mass loss rate of $2\times 10^{-5} M_\odot~{\rm yr}^{-1}$ \citep{cernicharo15}, at a projected separation of 
$100''$ ($1.8\times 10^{17}$ cm) from the star, the standard model H$_2$ density is 550 cm$^{-3}$.
Due to chemical reactions between gas species and on solid particle surfaces, and due to interaction with photons from 
the star and photons and cosmic rays from the interstellar medium, each gas species  has a distinct variation from
the bulk profile of equation \ref{massloss}. The most abundant parent molecule, H$_2$, and the solid particles, are formed 
in the dense environment close to the star where the gas is close to the condensation temperatures \citep[$>500$ K, ][]{wood19}.
The H$_2$ remains largely chemically inert through the portion of the envelope we are concerned with in this paper, spanning the resolvable envelope of the star out to where it merges with the ISM ($10^{15}$ to $10^{18}$ cm). 

Many molecular species are dissociated when exposed to the interstellar radiation field (ISRF), where they are exposed to  a
significant and sustained intensity of ultraviolet light. The situation is very similar to that
experienced by comet mass loss, where material is released from the comet surface then survives until
it is dissociated by sunlight or has a chemical reaction. We will borrow the standard terminology from comet studies and
call the molecules emerging from the hot inner envelope the parent molecules and their dissociation products the daughter molecules \citep{festou05}.
The abundant, readily observable, and chemically important parent molecule CO forms early in the outflow from the star and determines the  subsequent chemistry. If the stellar mass loss has more C than O, then nearly all the O is locked into CO and carbon chemistry determines the gas (e.g. hydrocarbons) and solid (e.g. polycyclic aromatic hydrocarbons, PAHs) composition. 
If the stellar mass loss has more O than C, then nearly all the C is locked in CO and the gas and oxygen chemistry determines the gas (e.h. H$_2$O) and solid (e.g. silicates) composition.
CO survives in the portion of the outflow where it is protected from UV photons in the ISRF by dust extinction and by self-shielding of overlapping absorption lines of the CO and H$_2$ molecules
\citep{mamon88, saberi19}. The hydrogen remains molecular in the region where CO is 
photodissociated, because H$_2$ continues to self-shield from dissociating photons in
that region.

\subsection{Basic models for carbon  in envelope}

The CO abundance relative to H$_2$ is characterized by the radius where it drops by half,
$r_{\rm phot}$, which is in the range $3\times 10^{15}$ to $10^{18}$ cm for a wide range of mass-loss rates and ISRF intensities \citep{saberi19}. 
For a mass-loss rate of $2\times 10^{-5} M_\odot~{\rm yr}^{-1}$,
{\bfred which applies for expansion velocity 15 km~s$^{-1}$ and [CO/H$_2$]$=8\times 10^{-4}$},
the photodissociation radius $r_{\rm phot}=3.6\times 10^{17}$ cm
\citep{mamon88}, corresponding to $200''$ projected separation from the star.
{\bfc The more recent calculation gives $r_{\rm phot}=3.0\times 10^{17}$ cm \citep{saberi19}.}
As a simple illustration, we use a `sharp' model where the CO/H$_2$ abundance is constant for
$r<r_{\rm phot}$, beyond which CO is photodissociated but H remains molecular. 
For a more accurate representation we use this exponential decrease that closely matches the theoretical models
for photodissociation:
\begin{equation}
    \frac{n_{\rm CO}}{n_{{\rm H}_2}} = 2 [{\rm C}/{\rm H}] e^{-{\rm ln}2 \left( r/r_{\rm phot}\right)^\alpha}.
\end{equation}
\citep{mamon88} where [C/H] is the elemental abundance of carbon relative to hydrogen and the factor of 2 is because there are two hydrogen atoms in each H$_2$ molecule.
For the abundance of carbon we consider two values: {\bfc  [CO/H$_2$]=$8\times 10^{-4}$ (so [C/H]=$4\times 10^{-4}$) }
that has been used in 
prior theoretical models for evolved star mass loss \citep{saberi19}, and
[C/H]=$2.1\times 10^{-4}$ is
based on analysis of photospheres of young, B-type stars \citep{nieva12}.

Upon dissociation of CO, the daughter products C and O are formed. 
{\bfc The O remains largely in neutral form in the diffuse ISM because its ionization potential is slightly higher than that of H, and the photons that would ionized O are absorbed by the more-abundance H instead.}
The C is rapidly ionized by ISRF photons above 11.26 eV, and the primary carbon daughter product of CO is C$^+$. 
For the density of atomic and ionized C, we assume it derives entirely from CO, so 
\begin{equation}
    \frac{n_{{\rm C}}}{n_{{\rm H}_2}} =  \left( 1 - \frac{n_{\rm CO}}{n_{{\rm H}_2}} \right) [{\rm CO}/{\rm H}_2]
\end{equation}
where the abundance of CO in the dense inner envelope is $[{\rm CO}/{\rm H}]\sim 8\times 10^{-4}$ from previous 
models  \citep{saberi19}.
Based only on cosmic abundances, the abundance [CO/H$_2$] = 2 [C/H] =$4.2\times 10^{-4}$ \citep{nieva12}. 
A higher [CO/H$_2$] than cosmic abundance 
means the stellar mass loss enriches the ISM, while a lower means it dilutes the C abundance.
The fraction of carbon that is ionized, at each distance from the star, was calculated  by balancing the recombination rate against 
the ionization rate from the ISRF extincted by the column density from that distance outward assuming an extinction per column density
as in the ISM.
To get a simple estimate of the C$^+$ column density, assuming a sharp CO/C$^+$ boundary at $r_{\rm phot}$, the column density of C$^+$ along the line
of sight through the star is
\begin{equation}
    N({\rm C}^+) = 2 n(r_{\rm phot}) r_{\rm phot} [{\rm CO}/{\rm H}_2]  \label{cpcol}
\end{equation}
where the initial factor of 2 is for the line of sight to the star and behind the star. 
For the standard model and a sharp dissociation of CO at $r_{\rm phot}$, the column density of 
C$^+$ on the line of sight through the star is $4.3\times 10^{16}$ cm$^{-2}$. 
For the various models considered in this paper, the column density
of C$^+$ remains within a factor of 2 of that value.

\begin{figure}
    \centering
    \includegraphics[width=4in]{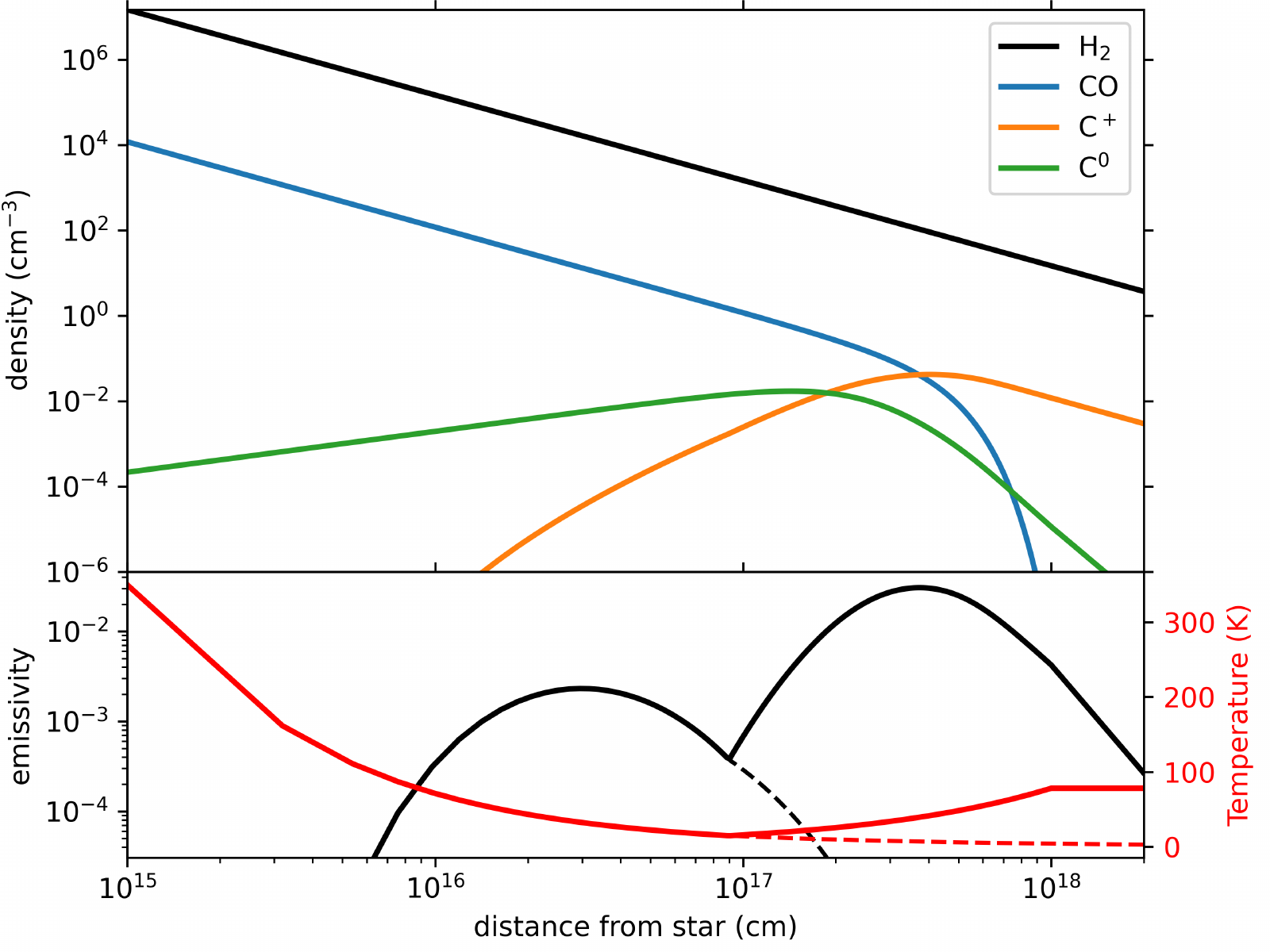}
    \caption{{\it (top panel}) Density of H$_2$, CO, C$^+$, and C$^0$ in the circumstellar envelope as a function of distance from the star. 
    The CO is dissociated by the ISRF leading to daughter products C$^0$ and
    C$^+$ in the outer envelope.  
    {\it (bottom panel}) Emissivity of [\ion{C}{2}] (left axis, 
    units $9\times 10^{-17}$ K~km~s$^{-1}$~cm$^{-1}$)
    and gas temperature (right axis) as a 
    function of distance to the star. Referring to Table~\ref{modtab},
    solid lines are for Model 0, in which the temperature increases in the outer envelope to meet that of the diffuse ISM, and dashed lines are for Model 0t, in which the temperature decreases to 10 K minimum throughout the outer envelope. Lacking the temperature increase in the outer envelope, Model 0t only
    produces [\ion{C}{2}] within $10^{17}$ cm of the star; on the other hand, Model 0 
    produces most [\ion{C}{2}] emission at $4\times 10^{17}$ cm from the star.
    }
    \label{fig:cpdens}
\end{figure}

The gas kinetic temperature is a balance between heating from the central star and from the external ISRF: there
are heat waves propagating from both directions. Cooling is via adiabatic expansion, plus molecular lines in the inner envelope and atomic fine structure lines in the outer envelope \citep{goldreich76,tielens85}. Models of the inner envelope have used an approximation 
$T = 14.6 (r_0/r)^\beta$ with $r_0=9\times 10^{16}$ cm, $\beta=0.72$ for $r<r_0$ and $\beta=0.54$ for $r>r_0$ \citep{kwan82,mamon88}. 
Measurements of gas temperature from CO lines in IRC+10216 yield temperatures of 20 K at $7\times 10^{16}$ cm and
15 K at $1\times 10^{17}$ cm \citep{guelin18}.
The CO molecules are therefore very cold when approaching the ISM, but exposure to the full ISRF both dissociates
the molecules and eventually heats the gas. 
In diffuse interstellar gas, observations of UV rotational lines of H$_2$ show temperatures of 78 K \citep{savage77}. 
{\bfc Cooling rates are shorter than the expansion time in the outer envelope. The material flowing out from IRC+10216 would reach comparable temperatures if it had 
an abundance of small carbon grains (which dominate photoelectric heating of the gas) comparable to that of the diffuse ISM. However, the abundance and charge of 
small carbon grains is not known. 
To bound temperature predictions in the shell beyond where the CO is dissociated, 
we use two illustrative models. In one model, we set a minimum temperature $T_{\rm min}$ for the gas.
In the other model, } we spliced the molecular gas temperature in the envelope  with a power-law that begins at $r_1=9\times 10^{16}$ cm and ends with 
$T=78$ K at $r_2=10^{18}$ cm to merge with the diffuse ISM. The values of $r_1$ and $r_2$ were chosen based
upon a static PDR model where we set the density  as per the standard model for IRC+10216, then illuminated with 
the ISRF from the outside, balancing photoelectric heating with cooling from fine structure and molecular line emission and where cooling by adiabatic expansion was neglected.

The most powerful spectral line from diffuse interstellar gas heated by the ISRF is the 157.7 $\mu$m 
{\bfc [\ion{C}{2}]} line. With an excited state 91 K above ground, this transition can be excited in the diffuse ISM, 
though it becomes exponentially more difficult to excite at low temperatures. 
We estimate the line brightness as a function of the temperature of the gas, density of H$_2$, and density of C$^+$,
following \citet{goldsmith12} to derive the emissivity then integrating through the envelope along lines
of sight separated from the star by 0 to 800$''$. 

Figure~\ref{fig:cpdens} shows the radial variation of the density, temperature,
and emissivity in the circumstellar envelope, for Models 0 and 0t.
The C$^0$ abundance never builds higher than 6\%
of gas-phase carbon, while CO is 100\% in the inner envelope and C$^+$ is 100\%
in the outer envelope. 
Model 0t has a cold outer envelope, and the  [\ion{C}{2}] emissivity peaks at around $3\times 10^{16}$ cm from the star.
Model 0, with the warmer outer envelope, has an additional 
 [\ion{C}{2}] emissivity peaks in the outer envelope near $R_{\rm phot}=3.65\times 10^{17}$ cm. 
For the other models, the relative amplitudes of the two 
peaks change. 
In Model 0t, the temperature of the outer envelope is allowed to decrease
as a power-law down to 10 K (dashed lines in Fig.~\ref{fig:cpdens}),  
and the outer peak is eliminated, so the [\ion{C}{2}] emission is significantly fainter and
closer to the star. If the temperature in the outer envelope is 20 K, the outer peak is also eliminated. If the temperature in
the outer envelope is 30 K, then the outer envelope dominates over the inner envelope.

\begin{figure}
    \centering
    \includegraphics[width=4in]{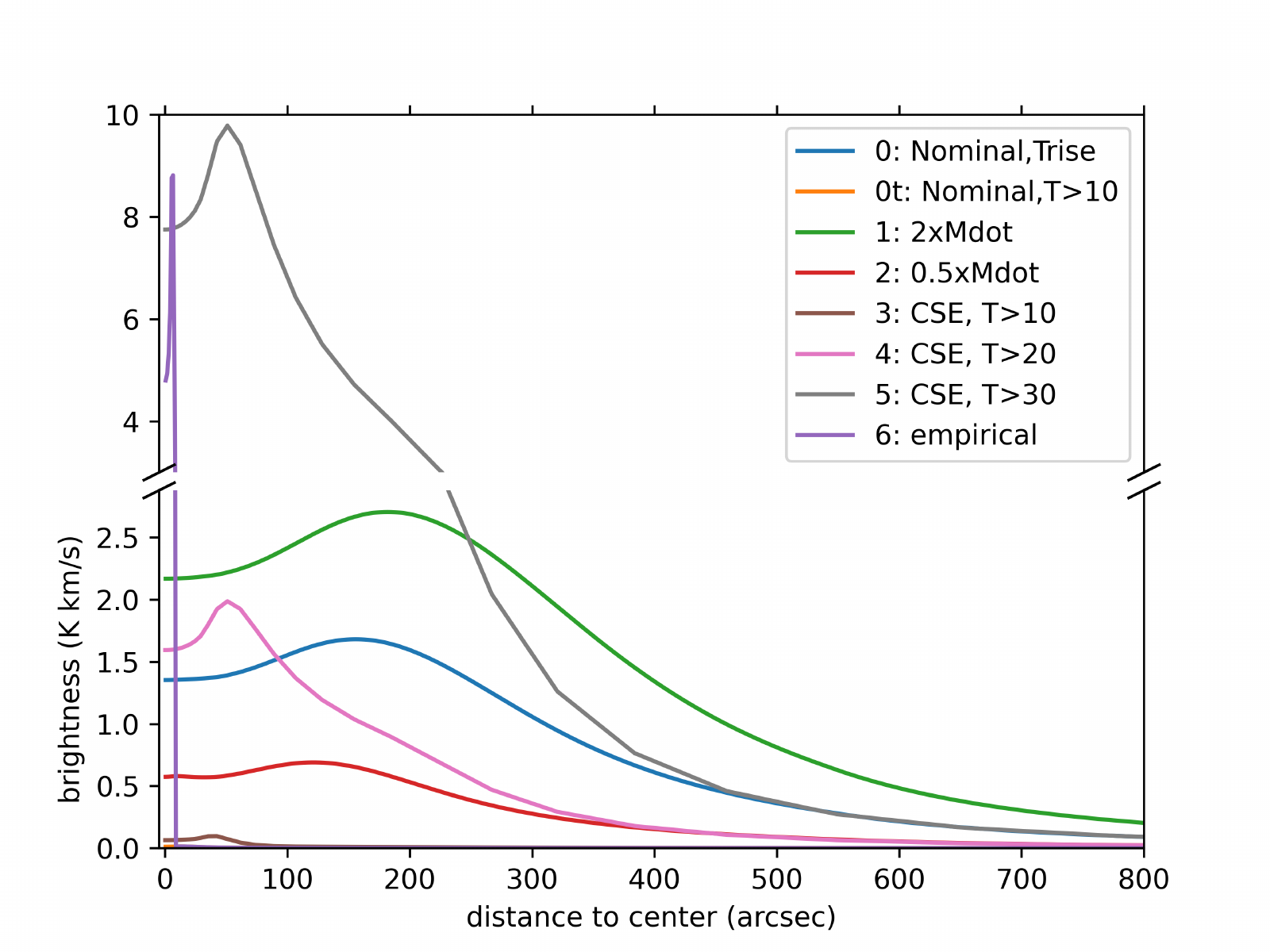}
    \caption{
    Predicted surface brightness of the [\ion{C}{2}] emission from IRC+10216 lines of sight with a range of separations from the star. The models are labeled according to their entries in Table~\ref{modtab}, where their parameters are listed. 
    Model 0 is a `nominal' prediction, being based upon prior models for CO dissociation, with temperature rising to meet that of H$_2$ in the diffuse ISM. Model 0t keeps
    the temperature at 10 K in the outer envelope and produces only a small integrated
    [\ion{C}{2}] brightness ($<0.015$ K).
     Models 1 and 2 have mass-loss rate
    increased and decreased by a factor of 2 (with corresponding shifts of $R_{\rm phot}$). 
    Models 3, 4, and 5 are from the CSE models (\S\ref{sec:PDR}), with different minimum temperatures. 
   Model 6 was empirically derived as discussed later in \S\ref{sec:discussion}, 
   adding for a shell of C$^+$ to match the observed brightness toward the star.
    }
    \label{fig:cpcalc}
\end{figure}

Figure~\ref{fig:cpcalc} shows the predicted brightness of  [\ion{C}{2}] for observations through the envelope at a range of projected separations
from the star. Predictions are shown for a set of models with parameters summarized in Table~\ref{modtab}. Model 0 is the baseline model, 
because it uses the photodissociation radius, CO density profile, and CO abundance from \citep{mamon88}, albeit with
the temperature modification described above, {\bfc rising to that of the diffuse ISM. Model 0t is the baseline model, but with the temperature decreasing to 10 K and remaining constant thereafter.
Models 1 and 2 have mass-loss rates that are higher and lower, respectively, than the baseline model; both models 1 and 2} use the rising temperature law, and the photodissociation radius was scaled by $\dot{M}^{-0.5}$
based upon the results of \citet{saberi19}.

\def\tnm{\tablenotemark}
\begin{deluxetable*}{ccccccc}
\tablewidth{0pt}
\tablecaption{Models for IRC+10216 [\ion{C}{2}]\label{modtab}} 
\tablehead{
\colhead{Model}  & \colhead{$R_{\rm phot}$} & \colhead{$\dot{M}$} & \colhead{Heating}  & 
 \colhead{$N({\rm C}^+)$} & \colhead{central $\int T dv$} & \colhead{max $\int T dv$} \\
 & \colhead{(cm)} & \colhead{($M_\odot$ yr$^{-1}$)} & \colhead{(note\tnm{a})} & \colhead{(cm)} & \colhead{(K km s$^{-1}$)} & \colhead{(K km s$^{-1}$)}
}
\startdata
0 & $ 3.65 \times 10^{17} $  & $ 2 \times 10^{-5} $ & rise78 & $3.3 \times 10^{16}$ &  1.4  &  1.7 \\
0t & $ 3.65 \times 10^{17} $  & $ 2 \times 10^{-5} $ & $T>10$ & $3.3 \times 10^{16}$ &  0.012  &  0.014 \\
1 & $ 4.47 \times 10^{17} $  & $ 3 \times 10^{-5} $ & rise78 & $4.1 \times 10^{16}$ &  2.2  &  2.7 \\
2 & $ 2.58 \times 10^{17} $  & $ 1 \times 10^{-5} $ & rise78 & $2.4 \times 10^{16}$ &  0.57  &  0.69 \\
3 & CSE (\S\ref{sec:PDR}) &   $ 2 \times 10^{-5} $ & $T >$ 10 & $6.8 \times 10^{16}$ &  0.064  &  0.096 \\
4 & CSE (\S\ref{sec:PDR}) &   $ 2 \times 10^{-5} $ & $T >$ 20 & $6.8 \times 10^{16}$ &  1.6  &  2.0 \\
5 & CSE (\S\ref{sec:PDR}) &   $ 2 \times 10^{-5} $ & $T >$ 30 & $6.8 \times 10^{16}$ &  7.7  &  9.8 \\
6 & note\tnm{b}  & $ 2 \times 10^{-5} $ & $T>10$ & $3.3 \times 10^{16}$ &  4.8  &  8.8 \\
\enddata
\tablenotetext{a}{Temperature distribution. All models begin at small $r$ with a
power-law decrease of $T(r)$. Models `$T>T_{\rm min}$' decrease to a minimum temperature $T_{\rm min}$. Models with `rise78' have 
the temperature decrease until $r=9\times 10^{16}$ cm then increase 
to 78 K at $10^{18}$ cm.}
\tablenotetext{b}{Model 6 is Model 0 with enhanced ionized carbon an inner shell from 1.5--2$\times 10^{16}$ cm from the star, as discussed in \S\ref{sec:discussion}.}
\end{deluxetable*}

Table~\ref{modtab} shows the model parameters and the resulting C$^+$ column density, brightness of the [\ion{C}{2}] line toward
the star, and maximum [\ion{C}{2}] brightness from the resolved envelope.
The predicted brightness is highest at a  separation from the star approximately 150--200$''$  
with a central depression due C$^+$ being absent from an inner, fully molecular region centered on the star.
The prediction is a thick shell, with central depression less than 50\%, and a tail extending out to large distances from the star. 
An inner peak is predicted for some models, due to the innermost C$^+$ warmed by the star. The presence or lack of an inner C$^+$ shell depends critically on the assumed temperature and the initial falloff of the CO density. The most critical factor determining the brightness of the [\ion{C}{2}] line is the mass loss rate. This can be readily understood from equation~\ref{cpcol}:  with the dependence of $n$ and $r_{\rm phot}$ on $\dot{M}$,
the column density of C$^+$ scales as $\dot{M}^{1.5}$. The second most important factor for determining the C$^+$ brightness is the abundance of CO in the envelope.

\subsection{Circumstellar envelope model for carbon in the envelope\label{sec:PDR}}

\begin{figure}
    \centering
    \includegraphics[width=4in]{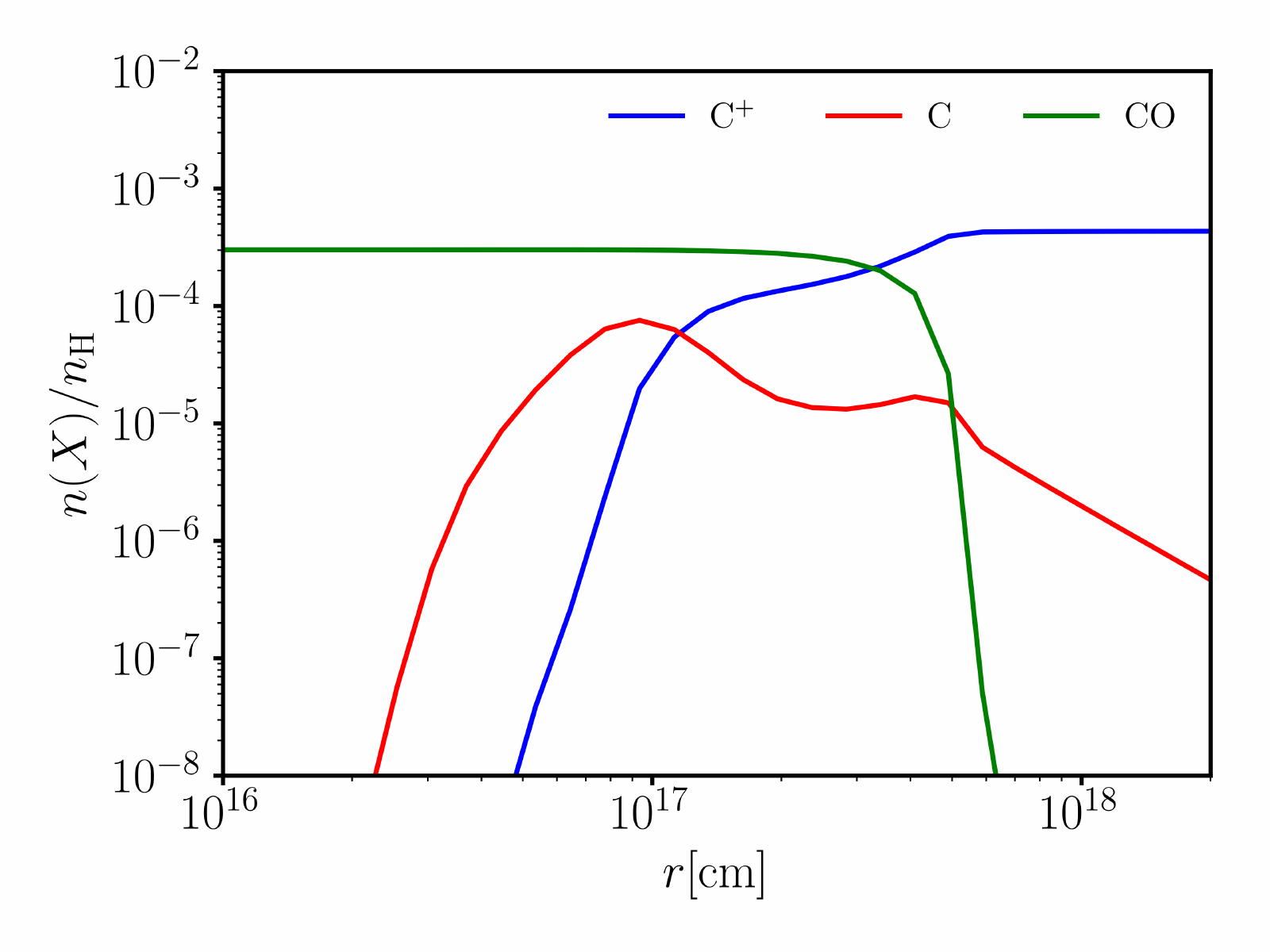}
    \caption{Carbon-bearing species abundances in the outer envelope of IRC+10216 from the CSE model of
    \S\ref{sec:PDR}. Gaseous C is assumed to be 75\% in CO molecules \citep[20\% in C$_2$H$_2$, 2.5\% in HCN;][]{agundez12}, with the rest of the C in solids, at the beginning of the outflow. The CO is photodissociated by the interstellar radiation field into C and O, and the C is rapidly ionized. The transition from CO to C$^+$ occurs at $3.5\times 10^{17}$ cm.
    }
    \label{fig:abun}
\end{figure}

In order to predict the [\ion{C}{2}] envelope brightness in more detail, 
we established a circumstellar envelope model similar to \citet{glassgold96} and \citet{millar00},
with the mass density
set per equation~\ref{massloss}, with inner radius $3\times 10^{14}$ cm.
The temperature profile was as per \citet{mamon88}, but with a threshold such that the temperature is never lower than a minimum $T_{\rm min}$.
This minimum temperature threshold allows us to explore the effects of additional sources of heating in the outer envelope that are important for predicting the brightness of the [\ion{C}{2}] line.
%were not taken into account in the computation of the temperature profile used in prior models.
The outer envelope temperature accounts for heating in the outer envelope that originates
from sources other than the CW Leo itself, for example photoelectric heating from UV-excited grains excited by the ISRF,
that can raise the temperature as 
discussed above. 
The chemistry is solved as a function of distance to the star 
\citep[][e.g.]{glassgold96}.
The chemical network is based on the {\tt kida.uva.2014} network \citep{wakelam15} which includes $\sim 500$ chemical species and $\sim 5000$ reactions. For gas-phase photoprocesses, we follow an approach similar to \citet{gorti04} to split  the FUV portion of the ISRF into nine energy bins from 0.74 to 13.6eV where the nine intervals were chosen to correspond to dominant gas photoabsorption thresholds. When available, destruction rates are computed by integrating photoabsorption, photodissociation, and photoionization cross sections over each energy bin to get the weighted cross sections for each species in each bin. The effect of self-shielding on the photodissociation rates is taken into account for H$_2$, CO, and N$_2$. H$_2$ self-shielding is treated following \citep{tielens85}. 
For CO and N$_2$, we use tabulated shielding functions that include the effects of self-shielding and the shielding provided by H and H$_2$ \citep{visser09,li13}. 
For absorption by  dust, we simply assume that the dust extinction scales with the total hydrogen column density and use 
${\rm A}_{\rm V} = 5.3 \times  10^{-22} N_{\rm H}$.
We use initial abundances from \citet{agundez12},
for which the total carbon and oxygen abundance are [C/H]= $4 \times 10^{-4}$
and [O/H]= $3 \times 10^{-4}$, leading to a [C/O] ratio of 1.3.

Figure~\ref{fig:abun} shows the computed abundance profile of C$+$, C$^0$ and CO 
as a function of the distance from the star. While CO is by far the dominant 
carrier of gas-phase carbon through most of the envelope, the daughter 
products begin appearing as close to the star as a significant abundance of C$^0$ at
$8\times 10^{16}$ cm then C$^+$ dominating outside $4\times 10^{17}$ cm, in accord
with the \citet{mamon88} and \citet{saberi19} calculations. 

Table~\ref{modtab} and
Figure~\ref{fig:cpcalc} (models 3, 4, and 5) shows the predicted line brightness for [\ion{C}{2}] emission for $T_{\rm min}=10$, 20, and 30 K, respectively.
The predicted brightness depends strongly upon the temperature profile, because 
the excited state is 91 K above ground: the predicted brightness changes by two
orders of magnitude for the modeled range of $T_{\rm min}$. 
The CSE models do not have a peak of
emission from the outer envelope (like Model 0 does, due to its assumed higher temperature in the envelope).
The CSE models have the [\ion{C}{2}] emission
more closely concentrated to the star, with a peak brightness for lines of sight
about $60''$ offset from the star due to gas at $10^{17}$ cm from the star.
The contrast between the central and peak brightness is about 25\%, so sensitive
observations with a resolution $<30''$ could in principle resolve the extended
envelope and detect limb-brightening.

\section{Observations of IRC +10216}

Expecting that evolved stars are  surrounded by halos of these daughter products, forming a thick shell centered
on the star, we designed an experiment to test the model on a well-known star. 

\subsection{SOFIA}

The Stratospheric Observatory for Infrared Astronomy \citep[SOFIA]{young12,temi18} enables measurement of
the [\ion{C}{2}] 157.7 $\mu$m line at an angular resolution of $14.1''$.
The line was observed with the
upgraded German Receiver at Terahertz Frequencies \citep[upGREAT][]{risacher16,risacher18}, which has a horns arranged in a
center-filled-hexagonal pattern and a backend capable of spectral resolution better than 1 km~s$^{-1}$. The low-frequency array was tuned to the [\ion{C}{2}] 
$^3P_1$--$3P_0$ line at 1900.536 GHz, 
and the high-frequency array was tuned to the [\ion{O}{1}]
$^3P_1$--$3P_2$ line at 4744.777 GHz.

The observing strategy was to take deep spectra, using the secondary mirror to chop N and S
of the envelope to cancel atmospheric emission. 
Each pointing of the array yielded 7 measurements of the [\ion{C}{2}] line at orthogonal linear polarizations, for 14 independent spectra. 
The array was offset from the star along a line straight east, with an offset of $96''$ 
between pointings, yielding a straight line of beams along the eastward direction
together with rows just N and S of that straight line offset N and S by $28''$. 
The first pointing was offset so that the closest spectrum to the star was $23.5''$ E
and the central row of the raster map was sampled approximately every $32''$ from there 
to $278''$ E of the star.
This set of pointings  covers the region where extended
C$^+$ is predicted to exist.
An additional pointing toward the GALEX-detected bowshock \citep{sahai10} 
was made along the same eastward line as the other
observations but offset 470$''$ E.

\def\tnm{\tablenotemark}
\begin{deluxetable*}{cccccc}
%\tabletypesize{\scriptsize} 
%\tablewidth{0pt}
%\tablenum{text}
%\tablecolumns{5}
\tablecaption{SOFIA Observing Log\label{obstab}} 
\tablehead{
\colhead{Date} & \colhead{Flight} & \colhead{Duration} & \colhead{Altitude} & \colhead{Position\tnm{a}} & \colhead{Chop}\\
\colhead{(UT)} & \colhead{} & \colhead{(min)} & \colhead{(feet)} & ($''$) & ($''$) \\
}
\startdata
04 Mar 2021 & 706 & 43 & 38,020 & 53, 149, 245 & $\pm 150''$, $\pm 200''$ NS \\
10 Mar 2021 & 709 & 30 & 40,500 & 53, 149, 245 & $\pm 200''$ NS\\
11 Mar 2021 & 709 & 20 & 43,000 & 53, 149, 245 & $\pm 200''$ NS\\
12 Mar 2021 & 710 & 68 & 43,000 & 470          & $300''$ E\\
\enddata
\tablenotetext{a}{East of IRC+10216 (CW Leo)}
\end{deluxetable*}

\begin{figure}
    \centering
    \includegraphics[width=4in]{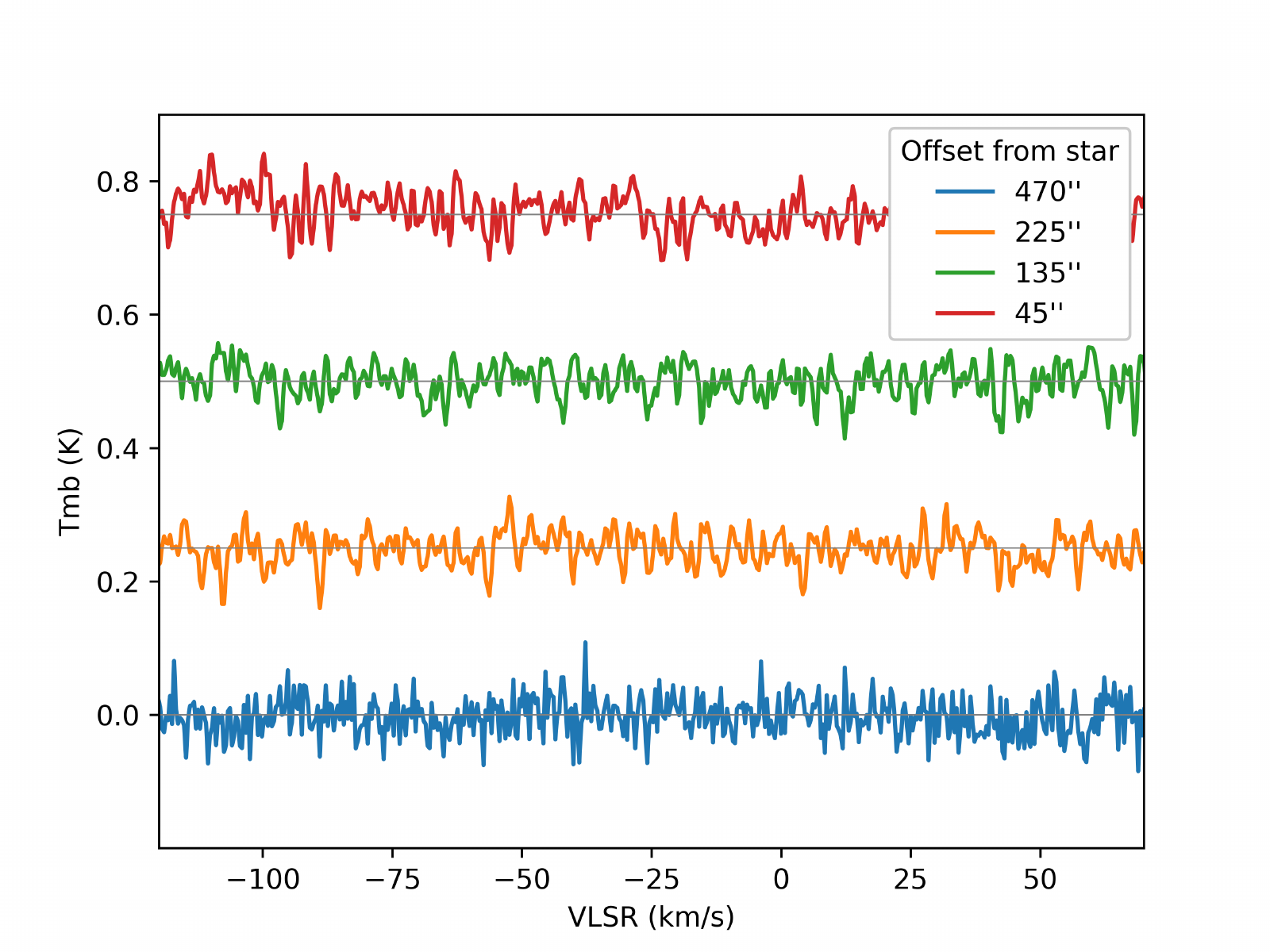}
    \caption{SOFIA/GREAT observations of IRC+10216 at four separations from the star, vertically offset for clarity.}
    \label{fig:obs}
\end{figure}

Observations took place during SOFIA's deployment to Cologne, Germany in March 2021
under program number 08\_0007.
Table~\ref{obstab} summarizes the observations.
On flight 709, two legs (one rising, one setting) at different altitudes were observed.
To minimize baseline structure (due to instabilities in the receiver and atmosphere)
a chop throw smaller than the total size
of the extended [\ion{C}{2}] cloud has to be utilized, so care must be taken in comparing
the models to the observations, as is done in \S\ref{sec:confront}.
For the raster map or three positions E of the star, a symmetric chop N and S was used.
For the bow shock position the secondary was chopped 300$''$ E only, to avoid chopping back onto the extended envelope.
In the end, each pointing in the raster map received 30 minutes of integration.

The calibration was done using the KOSMA atmospheric calibration software for SOFIA/GREAT 
\citep{guan12}. For the atmospheric 
opacity corrections, the precipitable water vapor column was obtained from a free fit to the atmospheric total power emission. The dry constituents were fixed to the standard model values. All receiver and system temperatures are on the single-sideband scale. 
The antenna temperature (T$^*_A$) was converted to T$_{\rm mb}$ using  beam efficiencies that were computed for each pixel and polarization ($\eta_{\mbox{MB}}= 0.63-0.69$). The main beam size is $14.1''$  for the LFA  array, and the error in the calibration is better than 20\%.

The data were reduced using the CLASS package from the GILDAS software\footnote{\url{https://www.iram.fr/IRAMFR/GILDAS/}}. 
A first-order baseline was removed from all spectra. One polarization from one of the horns (horn 5, vertical polarization) was excluded because its data were corrupted by interference. 
Figure~\ref{fig:obs} shows the [\ion{C}{2}] spectra, averaging the data in four distance ranges from the star. No lines were detected in the spectra with an rms brightness temperature of 0.03 K in spectra smoothed to 0.39 km~s$^{-1}$ resolution.
Because of the potential for self-chopping on the extended envelope, the data with $\pm 150''$
and $\pm 200''$ were reduced separately and found to be identical with no line detection.
The [\ion{C}{2}] line is predicted to be wider, but there is still no line
detected when smoothed to 3 km~s$^{-1}$ resolution, despite a low rms
brightness temperature of 0.015 K. For an upper limit to the [\ion{C}{2}] 
brightness, we use the rms from the spectra smoothed to 3 km~s$^{-1}$ times a
10 km~s$^{-1}$ linewidth based on other far-infrared
lines (see Herschel observations below), yielding $\int T dv <  0.15$ K~km~s$^{-1}$.
The [\ion{O}{1}] observations similarly showed no detection, with rms brightness temperature 0.1 K at 0.16 km~s$^{-1}$ resolution.

%Describe FIFI-LS spectra and reductions if they are used; also add coauthors NOT USING DUE TO CHOP
\def\extra{
FIFI-LS was pointed toward IRC+10216 as part of a calibration observation in 2021 May 25 for a 300 sec exposure. 
No line was evident in the FIFI-LS data, with an upper limit flux of $2\times 10^{-18}$ W~m$^{-2}$, which corresponds to an integrated line brightness of 5 K~km~s$^{-1}$. 
}

\subsection{Herschel}

\begin{figure}
    \centering
    \includegraphics[width=4in]{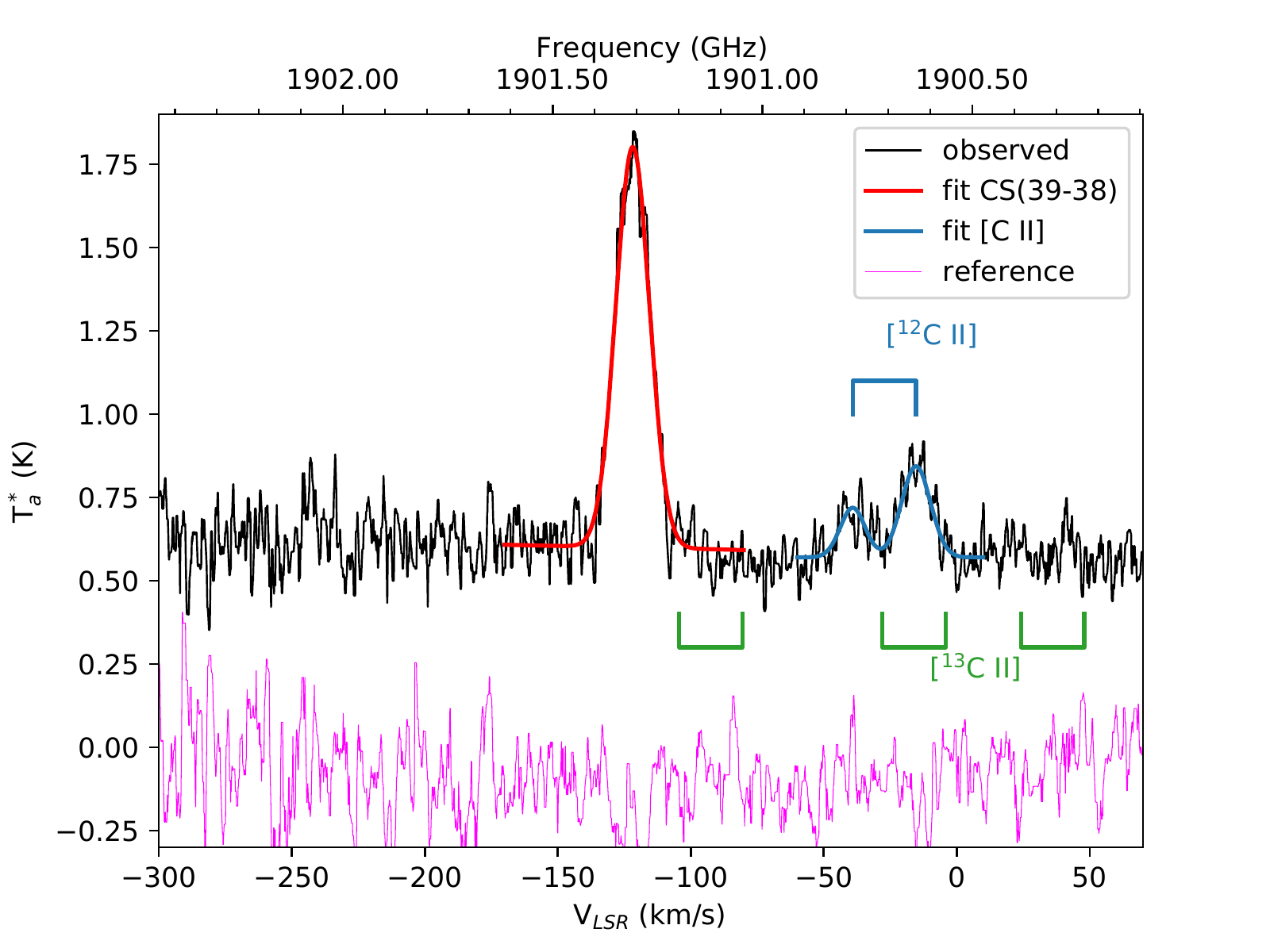}
    \caption{Herschel/HIFI observation of IRC+10216, with the lower horizontal axis being velocity relative to that of the rest 
    frame [\ion{C}{2}] line, and the upper axis being the rest frequency in the upper sideband. The bright line observed at  {\bfc USB frequency 1901.31 GHz (velocity -121 km~s$^{-1}$) is CS(39-38), which has rest frequency 1901.1510 GHz \citep{mueller05}}; a Gaussian profile fit is shown. 
    We identify the fainter two components at -15 and -39 km~s$^{-1}$ as [\ion{C}{2}],{\bfc  which has rest frequency 1900.5369 GHz}, and a model profile with the blue-shifted component absorbing the continuum is shown, as appropriate for an expanding shell with the star in the center (see text for details). 
    The magenta curve shows the reference position spectrum ($180''$ from the star). {\bfc Note the difference in baseline brightness between source and reference, which is due to dust emission from IRC+10216.}
    {\bfc Straight blue lines indicate the frequencies of the [$^{12}$C II] components, and straight green lines indicate the frequencies of the corresponding [$^{13}$C II] hyperfine transitions, which are not detected.}
    \label{fig:herschel}}
\end{figure}

The 1900.5369 GHz [\ion{C}{2}] frequency from IRC +10216 was observed by Herschel \citep{pilbratt10} with HIFI
\citep{degraauw10} as part of an early calibration observation {\bfc (2011 Oct 26),   and it was observed} with PACS \citep{poglitsch10} as part of a line survey (29 May 2011). The HIFI observation, at high spectral resolution, shows a bright line  
near---but not consistent with---the frequency of [\ion{C}{2}], making it impossible to separate detect the 
low spectral resolution of PACS. Therefore, we first investigate
the HIFI observation in detail. That short (875 sec duration) observation was taken using double beam switching, symmetrically offsetting the chopper $3'$ centered on the star with observation ID 1342231460. The data
were retrieved from the archive in $T_A^*$ units appropriate for a compact source \citep{shipman17}. We interpolated the two 
orthogonal polarization spectra to a common frequency grid and averaged them to create the final spectrum, shown
in Figure~\ref{fig:herschel}. {\bfc The Herschel telescope angular resolution is $11''$ at 157.7 $\mu$m.}
The reference spectrum ($3'$ from the star) was obtained from the Herschel archive \citep{teyssier16}, to evaluate possible contamination from extended emission as is often present due to the ISM in [\ion{C}{2}]. The reference
spectrum shows no such sign of emission lines, indicating the lines in the spectrum centered on the star are all from IRC+10216.
% TA* units, beam eff 0.56

HIFI spectra are double sideband, so a spectral line can appear from either lower or upper sidebands. The bright line near the center of the spectrum in Figure~\ref{fig:herschel} is at 1901.309 GHz (LSR frame). 
The line is likely CS(39-38) in the upper sideband with a -26.5 km~s$^{-1}$ LSR velocity that matches ground-based CO line observations \citep{cernicharo15,guelin18}, {\bfc based on the rest frequency in the Cologne Database for Molecular Spectroscopy  \citep[CDMS,][]{mueller05} of 1901.151 GHz. Archival Herschel observation also show the other CS rotational
lines as among the brightest in this frequency range; we verified 38-37, 37-36, 36-35, and 35-34.
(The frequency of CS(39-38) in the JPL molecular spectroscopy database is significantly different from CDMS, at 1900.732 GHz, but it is based on much older literature references. Based on the more modern references and the matching of all 4
other lines with CDMS, the line identification is of high confidence.) }
A Gaussian fit to this line has amplitude 1.12 K, FWHM 14.8 km~s$^{-1}$, and line integral 19 K~km~s$^{-1}$.
Fainter emission is at the correct frequency for [\ion{C}{2}] if it is also in the upper sideband and has two components, at
radial velocities -39.1 and -15.1 and km~s $^{-1}$. We interpret these components as the approaching and receding portions of the expanding outflow. 
The average of the central velocities of each of the two components is -27 km~s$^{-1}$, 
which is  consistent with the -26.5 km~s$^{-1}$ of the CS and CO lines. 
The separation between 
the velocities of the two components is 24 km~s$^{-1}$, also  consistent with the components
the approaching and receding portions of a shell expanding with velocity 15 km~s$^{-1}$
that has been determined from
prior studies of the star \citep[cf.]{cernicharo15}. The  flux of the [\ion{C}{2}] line is 5.1 K~km~s$^{-1}$
($1.0\times 10^{-16}$ W~m$^{-2}$).

The [\ion{C}{2}] line profile is noticeably asymmetric, with the blue-shifted component fainter than the red-shifted component.
This `P Cygni' like profile can arise if the approaching (blue-shifted) part of the outflow is in front of the star and absorbs the bright continuum from the dust envelope. From Akari observations, the total flux density of the star at 160 $\mu$m is 475 Jy \citep{ueta19}, which 
with Herschel HIFI the antenna temperature would convert to 0.87 K \citep{shipman17}. But at Herschel resolution, the shell is well resolved, so a lower brightness is observed, around 0.57 K if we look at frequencies away from the emission lines. If the blue-shifted component has a peak optical depth 0.25$^{+0.1}_{-0.05}$, then an excellent fit to the line profile is obtained (Fig.~\ref{fig:herschel}). {\bfc The integrated optical depth $\int \tau dv=3.0\pm0.4$ km~s$^{-1}$,
which requires a C$^+$ column density 
$(4\pm 0.6)\times 10^{17}$ cm$^{-2}$ \citep{goldsmith12}.
This is the column density integrated from the star to the observer through
the expanding shell.}

%  FREQ,           ERR, LGINT, DR,  ELO,    GUP, TAG, QNFMT,  QN',  QN" 
%  1900732.0307250.4494 -1.4321 2 1207.9033 79  44001 10139          38  
% 1948915.1976285.1413 -1.5328 2 1271.3048 81  44001 10140          39           

There is no evidence of extended [\ion{C}{2}] emission.
The extent of the [\ion{C}{2}] emission can be partly assessed from inspecting the HIFI reference spectrum at $180''$ from
the star, which shows no significant line emission, and for the SOFIA spectra showing no
emission $23''$ from the star, for which there is an even more strict upper limit. Therefore, the [\ion{C}{2}] is primarily compact. 
We inspected the PACS line observations to determine whether there is any evidence of extended emission near IRC+10216. The PACS spectral
imager has a $5\times 5$ array of spaxels of size $9.4''$ covering the $47''$ centered on the star. The spectral resolution of
240 km~s$^{-1}$ cannot separate the CS and [\ion{C}{2}] lines seen in the HIFI spectrum in Fig.~\ref{fig:herschel}. The CS(39-38) line arises from a much higher energy level (1800 K above ground) than that
of the [\ion{C}{2}] line (91 K above ground), so the former should arise only from a relatively hot region close to the star.
We used the PACS data to compare the spatial
distribution of those blended emission lines (at 157.69 $\mu$m) to that of a bright, nearby molecular line at 158.11 $\mu$m that is evident in the PACS spectrum and presumably also arising in the inner hot envelope. The emission from both features declines similarly with distance from the star, indicating no significant extended emission 
from [\ion{C}{2}] within $23''$ of the star.

{\bfc 
The isotopic [$^{13}$C II] fine structure comprises three hyperfine components with frequency offsets, converted to velocity offsets from
rest (i.e. where the [$^{12}$C II] line would appear) of -65.2 (F=1-0), +11.2 (F=2-1), and +63.2 (F=1-1) km~s$^{-1}$ \citep{cooksy86}. The relative intensities of the hyperfine components are 0.4, 1, and 0.2, respectively \citep{ossenkopf13}. Figure~\ref{fig:herschel} shows where the 
[$^{13}$C II] components would appear. 
The F=2-1 component would be blended with the [$^{12}$C II] line, and could conceivably contribute to the
apparent asymmetry of that line. There is a faint potential line at the location of the F=1-1 component, at $\simlt 0.1$ K and there is no evidence
of an F=1-0 component ($<0.08$ K). With the predicted branching ratios of the components, it is not possible for [$^{13}$C II] to be a significant contribution to the observed spectrum. Using the upper limit to F=1-0, the potential wiggle in the spectrum near the F=1-1 frequency would be too bright by a factor of 2.5. And even neglecting the lack of F=1-0, if the F=1-1 line were detected at 0.1 K, the central F=2-1 component should be 0.5 K which is brighter even than the [$^{12}$C II] line. Therefore, the observations indicate no [$^{13}$C II] emission, with a weak lower limit on the abundance ratio of
$^{12}$C/$^{13}$C$>2.3$, based on the ratio of observed [$^{12}$C II] to summed upper limits of the 
[$^{13}$C II] components. The lower limit is 
not surprising, given $^{12}$C/$^{13}$C is 89 in the solar photosphere \citep{clayton04}
and 68 in the ISM \citep{milam05}. In IRC+10216, the isotopic ratio $^{12}$CO/$^{13}$CO=40 \citep{guelin18}, so that $^{13}$CO dissociation may be expected to yield $^{12}$C/$^{13}$C closer to 40 for the daughter products. The lower limit from the present data does not approach these expected values.
}

\section{Confronting predictions with observations\label{sec:confront}}

\begin{figure}
    \centering
    \includegraphics[width=4in]{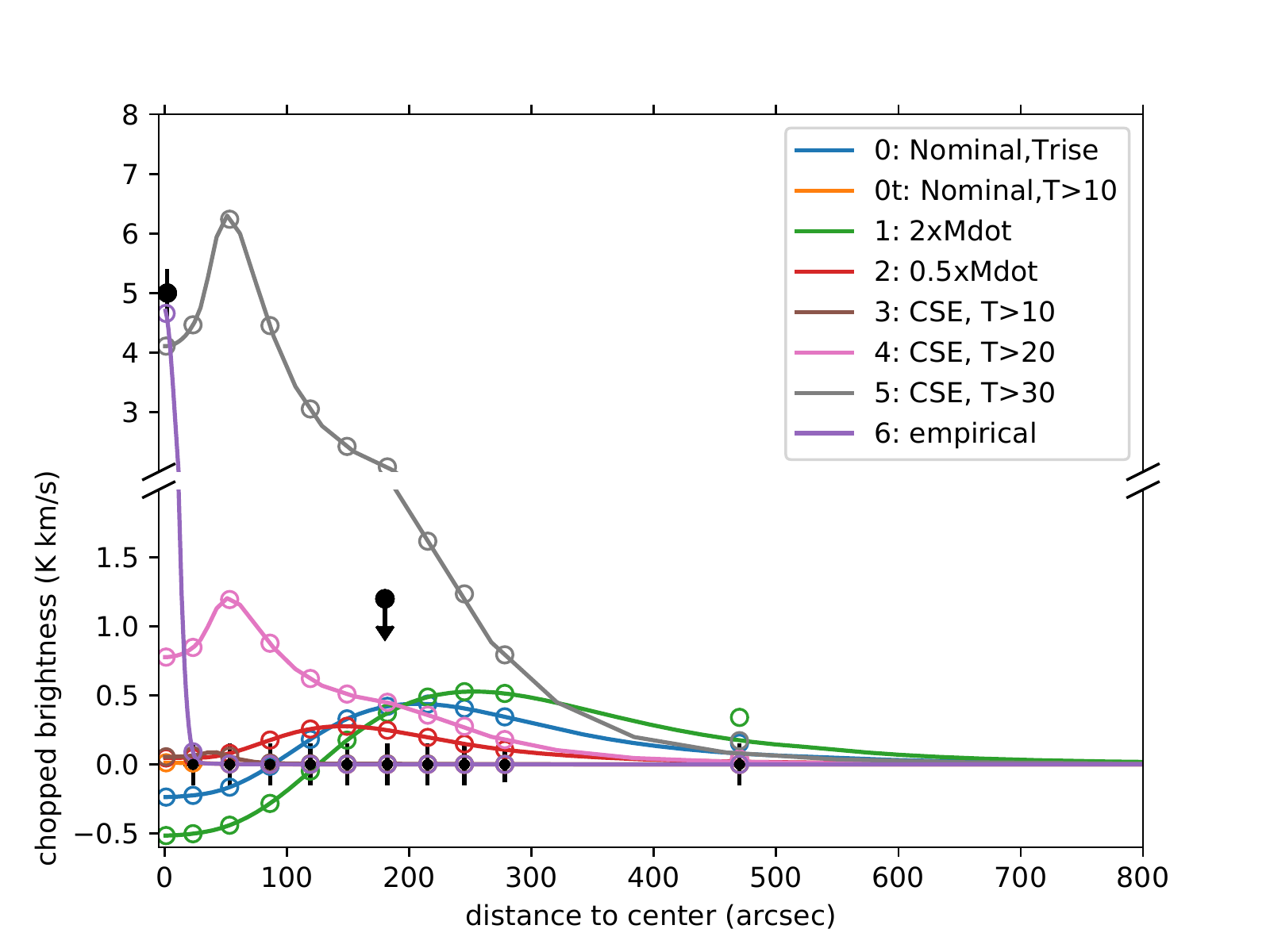}
    \caption{Comparison between the [\ion{C}{2}] line observations and models. Model parameters are listed in Table~\ref{modtab}. The model curves are the same as those in Fig.~\ref{fig:cpcalc}, but 
    virtually `chopped' using a $200''$ throw as in most of the SOFIA observations. Open circles are the models at the location of, and with simulated chop appropriate for, the specific SOFIA and Herschel observations. A negative chopped brightness means that the reference chop would have been brighter than the targeted position, so a spectral line of negative apparent brightness would appear with standard processing; such a feature would be just as readily evident in the data as a positive brightness.
    Black circles and bars denote the 95\% confidence upper limits from the SOFIA/GREAT observations.
    The black circle at zero  distance is the Herschel/HIFI detection toward the star, and the upper limit at $180''$  distance
     is from the Herschel/HIFI reference position spectrum. All of the models are ruled out with
     high confidence except the {\it post hoc} Model 3, which places a shell of
     C$^+$ in close to the star.
    }
    \label{fig:cpobs}
\end{figure}

The measured and upper-limit brightness of the [\ion{C}{2}] line are shown in Figure~\ref{fig:cpobs} as a function of distance from the star. Due to the large predicted size of the C$^+$ cloud around the star, and the differential, chopped nature of the observations with
chop throws of comparable size to the cloud, we generated `virtual chopped' models to allow a direct comparison between the model and the observations. We assumed the cloud is azimuthally symmetric, and for each modeled distance to the star, we subtracted the
brightness predicted to occur in the chop beam. The `virtual chopped' radial brightness profiles are shown in Figure~\ref{fig:cpobs} with the same
coloring scheme as the actual models from Figure~\ref{fig:cpcalc}. The predicted central brightness is suppressed in  the `chopped'
models of extended envelopes, because  the radial profile of the surface brightness is relatively flat out to the projected photodissociation
radius. 

In contrast to the models, the observations show (1) a detection toward the star, brighter than predicted by any of the models, and (2) upper limits on extended emission, fainter than predicted models in which the outer envelope is warmer than 10 K.
Further, comparing to the C$^+$ column density predictions in Table~\ref{modtab} to the column density derived toward the star
from the Herschel/HFI spectrum, none of the models have high enough column density of C$^+$.

Thus, while the extended envelope is not detected, the inner envelope 
requires more C$^+$ than predicted. It is possible to explain the lack of [\ion{C}{2}]
emission from the outer envelope as being due to low temperatures there, such as in Model 0t. While this did not 
agree with our expectation that the ISRF would heat the envelope, the balance of heating and cooling depends upon the nature of the macromolecules and solids that contain most of the heavy elements in the envelope. 
If the photoelectric heating is suppressed in the envelope,
compared to the diffuse ISM, then the temperature could be low
in the envelope, even for densities comparable to the diffuse ISM. 
Also, if the mass-loss is non-steady such that the envelope is not smoothly filled, then the
local volume density could be higher than large-scale averages, which would enhance the
cooling due to collisionally-excited lines from the gas. 
%Azimuthal asymmetries or patchiness, if present, would allow for deeper penetration of UV light from the ISRF.
Furthermore, the gas is not necessarily in thermal equilibrium, and the adiabatic cooling during its expansion could lower the temperature enough that the $^3{\rm P}_1$ level of
C$^+$ is not excited.
Any of these effects (decreased heating, enhanced cooling, and adiabatic expansion) could keep the gas cold and suppress the [\ion{C}{2}] emission from the envelope. 
But the  emission and absorption on the line of sight to the star remain unexplained 
by the {\it a priori} models.

\section{How to explain the gas-phase carbon distribution\label{sec:discussion}}

In order to explain the [\ion{C}{2}] emission and absorption toward the star, we introduce
a shell of C$^+$
at a distance from the star $r$, with a thickness $r/2$, and we use observational results to constrain $r$.
The atomic gas in the shell contains a fraction $f_{\rm C}$ of the gas-phase carbon available at $r$.
Then the column density 
\begin{equation}
    N({\rm C}^+) = n(r) r f_{\rm C}
\end{equation}
where $n(r)$ is from equation~\ref{massloss}. 
The material must not extend far enough that it would be resolved and detected in the
SOFIA spectra, which limits $r<3\times 10^{16}$ cm.
Using the observed column density from the blue-shifted absorption in the Herschel spectrum, the shell could occur 
at $r=1.2\times 10^{16}$ cm, if it contains $f=60$\% of the gas-phase carbon there. 
This close to the star, external ISRF heating is negligible, and the gas temperature 
inferred from the astrochemistry models and observations \citep{saberi19,cernicharo15}
is 62 K. {\bfc The upper level of the [\ion{C}{2}] line is readily excited in this inner envelope. 
The density must fall off very quickly in order to remain consistent
with the SOFIA and Herschel upper limits 
in the outer envelope.}

The [\ion{C}{2}] observational result is qualitatively similar to that obtained for [\ion{C}{1}]
by \citet{keene93}, who detected optically-thin emission in the 492 GHz line from the
$^3$P$_1$-$^3$P$_0$ transition.
They fit the spatial and spectra distribution of the emission with shells at 
$2.7\times 10^{16}$ cm and $8.2\times 10^{16}$ cm
radius, 
with a column density in the upper level inferred to be 
$N({\rm C}^0, ^3P_1)=7\times 10^{15}$ cm$^{-2}$ and $4\times 10^{15}$ cm$^{-2}$, respectively.
The inferred locations of the C$^0$ shells is outside those we inferred for the C$^+$.
For a concordance, the C$^+$ would need to be a bit further out, and $f_{\rm C}$ lower,
which is only marginally allowed given the upper limit to the [\ion{C}{2}] emitting region size. 
Using the same temperature law as described above (heating only from the star), the inferred inner and outer
C$^0$ shells are at 35 K at 16 K, respectively. 
The upper level of the 492 GHz fine structure line is readily excited at such temperatures,
and the  {\bfc total (not just upper level)} C$^0$ column densities of the shells are 
$1.2\times 10^{16}$ cm$^{-2}$ and $1.0\times 10^{16}$ cm$^{-2}$.
The column density of C$^+$ is much higher than that of C$^0$, suggesting that
the ionized carbon in the inner envelope rapidly recombines and is incorporated into molecules.

\begin{figure}
    \centering
    \includegraphics[width=4in]{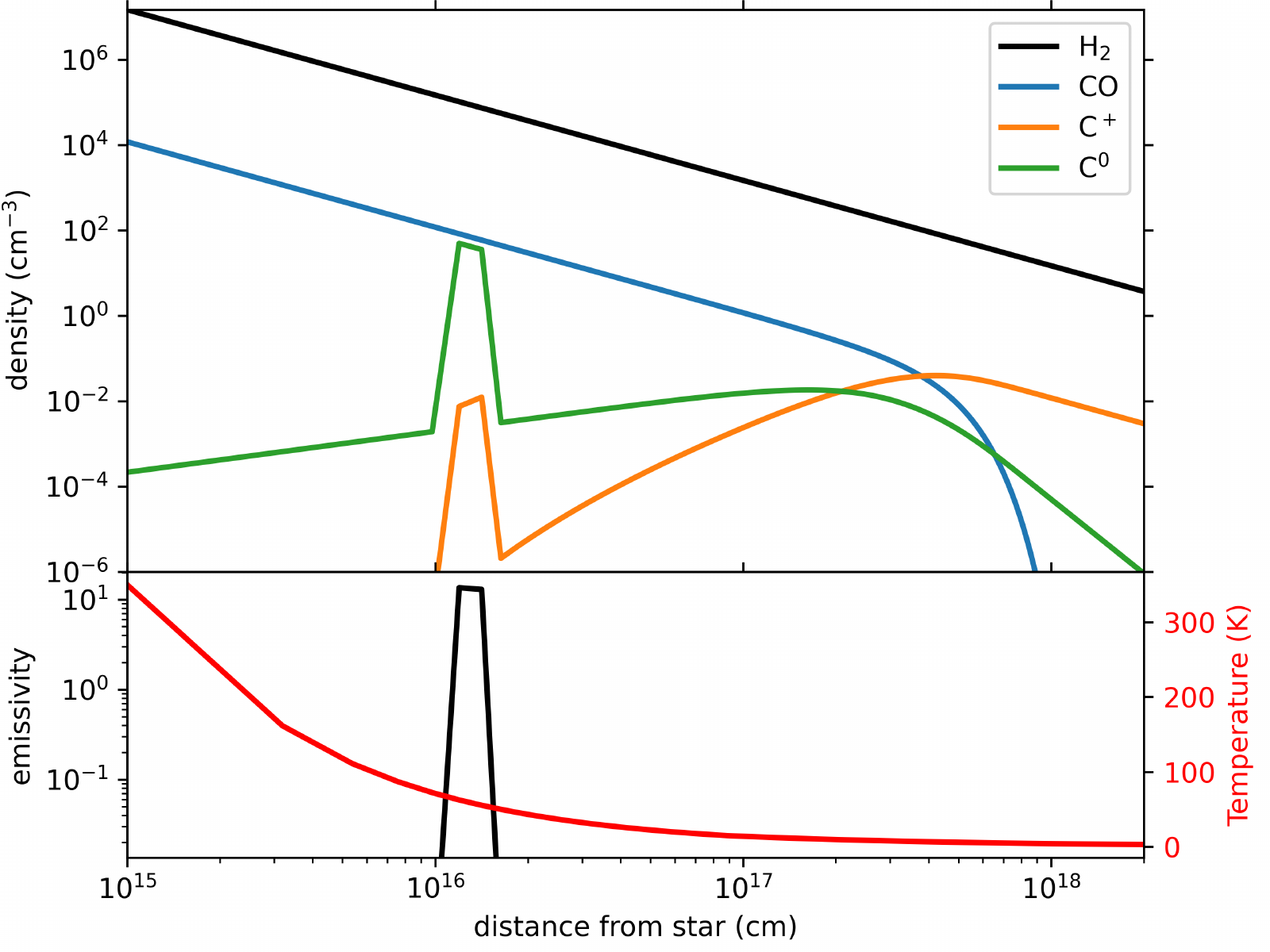}
    \caption{
    \def\extra{\bf THIS FIGURE IS OPTIONAL, NOT SURE IF IT IS FULLY NECESSARY.} 
    Densities and emissivity for an empirical model that combines a cold envelope of daughter products with an additional inner shell of C$^+$ tuned to match the [\ion{C}{2}] detection toward
    the star. Same as Figure~\ref{fig:cpdens} but for the empirical model discussed
    in \S\ref{sec:discussion} and listed as Model 6 in Table~\ref{modtab}.
     }
    \label{fig:cpden2s}
\end{figure}

The amount of free gas-phase carbon (i.e. not  in molecules or dust) required to explain the
observed [\ion{C}{2}] absorption is about $2\times 10^{-6} M_\odot$ 
(for a column density $4\times 10^{17}$ cm$^{-2}$ in a region of radius $1.2\times 10^{16}$ cm),
which is a small fraction ($<0.1$\%) of the total yield of C from a solar metalicity
AGB star with main-sequence mass 1.5--4.5 $M_\odot$ \citep{karakas10}. 
The column density we derive for C$^+$ is 9 times higher than predicted by the
detailed chemical model of \citet{millar00}. Similarly, the column density
derived of C$^0$ is 2.2 times higher than the model. 
In the warm, dense environment near the star, gas-phase carbon is rapidly consumed
into molecules like CO and C$_2$H$_2$, within timescales of order 1 yr for gas
within $10^{15}$ cm of the star. However, if any carbon were liberated to the gas
phase in the outer envelope, then it can survive much longer, due to the rapid $r^{-2}$ 
decrease in gas density and $n^2$ nature of chemical reactions. For example, at
$10^{16}$ cm from the star, chemical timescales are $>10^3$ yr. For comparison,
the dynamical timescale of gas expansion is $r/v=210$ yr at that same distance from
the star. Therefore, carbon formed from photodissociation at $\sim 10^{16}$ cm will
flow together with the molecules outward to the rarefied outer parts of the envelope.

There was no sign of [\ion{C}{2}] or [\ion{O}{1}] emission from the bow shock that is $8.6\times 10^{17}$ cm (470$''$) from the star due to its motion through the surrounding ISM
\citep{sahai10,ladjal10}. The actual velocity is not well known, having been inferred from 
the location of the termination shock and assumption about the ISM density (which is unknown). For example if the motion is 90 km~s$^{-1}$, the ISM will experience a 44 km~s$^{-1}$ J-type shock and the outer envelope will experience a ~7 km~s$^{-1}$
wind termination shock. The upper limit from SOFIA of 0.15 K km~s$^{-1}$ corresponds to
a surface brightness $10^{-6}$ erg~s~cm$^{-2}$~sr$^{-1}$. For a radiative shock, the 
total energy loss rate is  $\rho v^3/8\pi \sim 10^{-5} n_0 v_{45}^3$ erg~s~cm$^{-2}$~sr$^{-1}$, where $n_0$ is the surrounding
ISM density (cm$^{-3}$), and $v_{45}$ is the star's velocity divided by 45 km~s$^{-1}$.
The [\ion{C}{2}] upper limit constrains the pre-shock ISM density to
$n_0 < 1$ cm$^{-3}$ if 10\% of the radiative losses are in the [\ion{C}{2}] line.
Limiting the power of the shock also limits its UV production. The GALEX FUV 
intensity is interpreted as fluorescent H$_2$ rather than shocks or dust
scattering \citep{sahai10}, so there is only weak UV radiation from the bow-shock,
which is unlikely to significantly affect the circumstellar envelope except in the immediate vicinity of the bow shock itself. 
Estimating the envelope density at $8.6\times 10^{17}$ cm from the star
as 20 cm$^{-3}$ and the reverse (termination) shock velocity as 7 km~s$^{-1}$, the power in the reverse 
shock is about 10 times lower than the bow shock, with negligible UV production.

\section{Conclusions\label{sec:conclusion}}

For the standard model of mass loss considered in this paper, CO is a parent molecule, with C being a daughter product and C$^+$ a quickly-formed grand-daughter. 
We do not detect emission from the C$^+$ that should exist at the outer edge of the CO 
envelope. The strict upper limits require that the carbon gas is injected into the ISM is 
very cold ($\simlt 20$ K), either due to its rapid adiabatic expansion or due to a lack of the heating mechanisms that apply to average ISM gas (where photoelectrons from small PAH
dominate heating).
If there is another C-bearing parent molecule, with a very short lifetime to  dissociation
by interstellar radiation field photons in the outflow, then it could in principle explain an inner shell of carbon. 
Hypothetically, such a parent would be rich in C and susceptible to dissociation by longer-wavelength photons
that penetrate deeper into the envelope. The photons from the star could also contribute, forming a two-sided PDR with
the star and its thick dust shell on the inside and the ISRF on the outside. The presence of a remarkable wealth of complex
molecules in the IRC+10216 including CH$_2$CHCN, CH$_2$CN, CH$_3$CCH, and H$_2$CS \citep{agundez10}, with {\bfred prominent} C-bearing species
in the inner envelope within $2\times 10^{15}$ cm of the star including C$_2$H$_2$ and HCN at abundances within an order of magnitude of CO \citep{agundez12}, encourages such speculation.
The atomic carbon in the inner envelope has been suggested to be dissociation products of
C$_2$H$_2$ and CO, though the locations of those shells is much closer to the star than
expected for ISRF photodissociation of those molecules \citep{keene93}.

Carbon formed in the inner envelope is rapidly incorporated back into hydrocarbons, so
maintaining a large column density of atomic and ionized carbon
requires an additional source of dissociating photons.
The average photospheric emission of the central star of IRC+10216, CW Leo, has
insufficient blue to ultraviolet light. The source of the additional dissociation 
could be a companion star, the existence of which has already been inferred from
the geometry of shells of mass loss \citep{cernicharo15}, or shocks in the inner envelope.
The bow shock would primarily affect the outer envelope, which appears unperturbed, being
primarily molecular and having roughly spherical geometry, 
throughout the CO emission region. The SOFIA and Herschel observations can only be
explained with ionization well inside the CO emitting region.
Visible and ultraviolet light from accretion shocks of material onto a companion star
are a potential source of dissociating photons.
A white dwarf companion would produce enough dissociating and ionizing photons to
significantly alter the circumstellar chemistry and lead to C$^+$ deep in the envelope
\citep{millar20}.
Atomic carbon near evolved stars has also been detected for 
$\alpha$ Ori \citep{huggins94} and $o$ Cet \citep{saberi18} and interpreted as
potentially due to presence of UV radiation from a chromosphere or accretion shocks.
{\bfc In R Scl, atomic carbon was detected from a detached shell, due to dissociation of molecules other than CO \citep{oloffson15}.}
The  column density of atomic and ionized carbon in the inner envelope of IRC+10216 is estimated to be
$4\times 10^{17}$ cm$^{-2}$. For comparison, for theoretical models of outflows, the
total column density of carbon-bearing gas species (primarily CO, C$_2$H$_2$, and HCN) is $2\times 10^{18}$ cm$^{-2}$ \citep{millar00}, so the additional carbon implied by the new observations is
a significant but not dominant constituent of the total envelope.

\def\extra{
include prediction for other stars, higher mass loss
relate again deathstar, showing the CO bump and C+ halo together, complementarity to ALMA survey of mass-loss \citep{ramstedt20}}

\acknowledgements  
Based in part on observations made with the NASA/DLR Stratospheric Observatory for Infrared Astronomy (SOFIA). SOFIA is jointly operated by the Universities Space Research Association, Inc. (USRA), under NASA contract NNA17BF53C, and the Deutsches SOFIA Institut (DSI) under DLR contract 50 OK 0901 to the University of Stuttgart. 
Financial support for this work was provided by NASA (through award \#08-0007) issued by USRA.
Herschel is an ESA space observatory with science instruments provided by European-led Principal Investigator consortia and with important participation from NASA. 
This work made use of the open-source python packages {\tt matplotlib} \citep{matplotlib} and {\tt astropy} \citep{astropy}.

\facilities{SOFIA,Herschel}
%\clearpage

\bibliography{wtrbib}

\begin{thebibliography}{}
\expandafter\ifx\csname natexlab\endcsname\relax\def\natexlab#1{#1}\fi

\bibitem[{{Ag{\'u}ndez} {et~al.}(2012){Ag{\'u}ndez}, {Fonfr{\'\i}a},
  {Cernicharo}, {Kahane}, {Daniel}, \& {Gu{\'e}lin}}]{agundez12}
{Ag{\'u}ndez}, M., {Fonfr{\'\i}a}, J.~P., {Cernicharo}, J., {et~al.} 2012,
  \aap, 543, A48

\bibitem[{{Ag{\'u}ndez} {et~al.}(2008){Ag{\'u}ndez}, {Fonfr{\'\i}a},
  {Cernicharo}, {Pardo}, \& {Gu{\'e}lin}}]{agundez10}
{Ag{\'u}ndez}, M., {Fonfr{\'\i}a}, J.~P., {Cernicharo}, J., {Pardo}, J.~R., \&
  {Gu{\'e}lin}, M. 2008, \aap, 479, 493

\bibitem[{{Astropy Collaboration} {et~al.}(2018){Astropy Collaboration},
  {Price-Whelan}, {Sip{\H{o}}cz}, {G{\"u}nther}, {Lim}, {Crawford}, {Conseil},
  {Shupe}, {Craig}, {Dencheva}, {Ginsburg}, {Vand erPlas}, {Bradley},
  {P{\'e}rez-Su{\'a}rez}, {de Val-Borro}, {Aldcroft}, {Cruz}, {Robitaille},
  {Tollerud}, {Ardelean}, {Babej}, {Bach}, {Bachetti}, {Bakanov}, {Bamford},
  {Barentsen}, {Barmby}, {Baumbach}, {Berry}, {Biscani}, {Boquien}, {Bostroem},
  {Bouma}, {Brammer}, {Bray}, {Breytenbach}, {Buddelmeijer}, {Burke},
  {Calderone}, {Cano Rodr{\'\i}guez}, {Cara}, {Cardoso}, {Cheedella}, {Copin},
  {Corrales}, {Crichton}, {D'Avella}, {Deil}, {Depagne}, {Dietrich}, {Donath},
  {Droettboom}, {Earl}, {Erben}, {Fabbro}, {Ferreira}, {Finethy}, {Fox},
  {Garrison}, {Gibbons}, {Goldstein}, {Gommers}, {Greco}, {Greenfield},
  {Groener}, {Grollier}, {Hagen}, {Hirst}, {Homeier}, {Horton}, {Hosseinzadeh},
  {Hu}, {Hunkeler}, {Ivezi{\'c}}, {Jain}, {Jenness}, {Kanarek}, {Kendrew},
  {Kern}, {Kerzendorf}, {Khvalko}, {King}, {Kirkby}, {Kulkarni}, {Kumar},
  {Lee}, {Lenz}, {Littlefair}, {Ma}, {Macleod}, {Mastropietro}, {McCully},
  {Montagnac}, {Morris}, {Mueller}, {Mumford}, {Muna}, {Murphy}, {Nelson},
  {Nguyen}, {Ninan}, {N{\"o}the}, {Ogaz}, {Oh}, {Parejko}, {Parley}, {Pascual},
  {Patil}, {Patil}, {Plunkett}, {Prochaska}, {Rastogi}, {Reddy Janga},
  {Sabater}, {Sakurikar}, {Seifert}, {Sherbert}, {Sherwood-Taylor}, {Shih},
  {Sick}, {Silbiger}, {Singanamalla}, {Singer}, {Sladen}, {Sooley},
  {Sornarajah}, {Streicher}, {Teuben}, {Thomas}, {Tremblay}, {Turner},
  {Terr{\'o}n}, {van Kerkwijk}, {de la Vega}, {Watkins}, {Weaver}, {Whitmore},
  {Woillez}, {Zabalza}, \& {Astropy Contributors}}]{astropy}
{Astropy Collaboration}, {Price-Whelan}, A.~M., {Sip{\H{o}}cz}, B.~M., {et~al.}
  2018, \aj, 156, 123

\bibitem[{{Becklin} {et~al.}(1969){Becklin}, {Frogel}, {Hyland}, {Kristian}, \&
  {Neugebauer}}]{becklin69}
{Becklin}, E.~E., {Frogel}, J.~A., {Hyland}, A.~R., {Kristian}, J., \&
  {Neugebauer}, G. 1969, \apjl, 158, L133

\bibitem[{{Busso} {et~al.}(1999){Busso}, {Gallino}, \& {Wasserburg}}]{busso99}
{Busso}, M., {Gallino}, R., \& {Wasserburg}, G.~J. 1999, \araa, 37, 239

\bibitem[{{Cernicharo} {et~al.}(2015){Cernicharo}, {Marcelino}, {Ag{\'u}ndez},
  \& {Gu{\'e}lin}}]{cernicharo15}
{Cernicharo}, J., {Marcelino}, N., {Ag{\'u}ndez}, M., \& {Gu{\'e}lin}, M. 2015,
  \aap, 575, A91

\bibitem[{{Clayton} \& {Nittler}(2004)}]{clayton04}
{Clayton}, D.~D., \& {Nittler}, L.~R. 2004, \araa, 42, 39

\bibitem[{{Cooksy} {et~al.}(1986){Cooksy}, {Blake}, \& {Saykally}}]{cooksy86}
{Cooksy}, A.~L., {Blake}, G.~A., \& {Saykally}, R.~J. 1986, \apjl, 305, L89

\bibitem[{{de Graauw} {et~al.}(2010){de Graauw}, {Helmich}, {Phillips},
  {Stutzki}, {Caux}, {Whyborn}, {Dieleman}, {Roelfsema}, {Aarts}, {Assendorp},
  {Bachiller}, {Baechtold}, {Barcia}, {Beintema}, {Belitsky}, {Benz}, {Bieber},
  {Boogert}, {Borys}, {Bumble}, {Ca{\"\i}s}, {Caris}, {Cerulli-Irelli},
  {Chattopadhyay}, {Cherednichenko}, {Ciechanowicz}, {Coeur-Joly}, {Comito},
  {Cros}, {de Jonge}, {de Lange}, {Delforges}, {Delorme}, {den Boggende},
  {Desbat}, {Diez-Gonz{\'a}lez}, {di Giorgio}, {Dubbeldam}, {Edwards},
  {Eggens}, {Erickson}, {Evers}, {Fich}, {Finn}, {Franke}, {Gaier}, {Gal},
  {Gao}, {Gallego}, {Gauffre}, {Gill}, {Glenz}, {Golstein}, {Goulooze},
  {Gunsing}, {G{\"u}sten}, {Hartogh}, {Hatch}, {Higgins}, {Honingh}, {Huisman},
  {Jackson}, {Jacobs}, {Jacobs}, {Jarchow}, {Javadi}, {Jellema}, {Justen},
  {Karpov}, {Kasemann}, {Kawamura}, {Keizer}, {Kester}, {Klapwijk}, {Klein},
  {Kollberg}, {Kooi}, {Kooiman}, {Kopf}, {Krause}, {Krieg}, {Kramer},
  {Kruizenga}, {Kuhn}, {Laauwen}, {Lai}, {Larsson}, {Leduc}, {Leinz}, {Lin},
  {Liseau}, {Liu}, {Loose}, {L{\'o}pez-Fernandez}, {Lord}, {Luinge}, {Marston},
  {Mart{\'\i}n-Pintado}, {Maestrini}, {Maiwald}, {McCoey}, {Mehdi}, {Megej},
  {Melchior}, {Meinsma}, {Merkel}, {Michalska}, {Monstein}, {Moratschke},
  {Morris}, {Muller}, {Murphy}, {Naber}, {Natale}, {Nowosielski}, {Nuzzolo},
  {Olberg}, {Olbrich}, {Orfei}, {Orleanski}, {Ossenkopf}, {Peacock}, {Pearson},
  {Peron}, {Phillip-May}, {Piazzo}, {Planesas}, {Rataj}, {Ravera}, {Risacher},
  {Salez}, {Samoska}, {Saraceno}, {Schieder}, {Schlecht}, {Schl{\"o}der},
  {Schm{\"u}lling}, {Schultz}, {Schuster}, {Siebertz}, {Smit}, {Szczerba},
  {Shipman}, {Steinmetz}, {Stern}, {Stokroos}, {Teipen}, {Teyssier}, {Tils},
  {Trappe}, {van Baaren}, {van Leeuwen}, {van de Stadt}, {Visser}, {Wildeman},
  {Wafelbakker}, {Ward}, {Wesselius}, {Wild}, {Wulff}, {Wunsch}, {Tielens},
  {Zaal}, {Zirath}, {Zmuidzinas}, \& {Zwart}}]{degraauw10}
{de Graauw}, T., {Helmich}, F.~P., {Phillips}, T.~G., {et~al.} 2010, \aap, 518,
  L6

\bibitem[{{Decin} {et~al.}(2012){Decin}, {Cox}, {Royer}, {Van Marle},
  {Vandenbussche}, {Ladjal}, {Kerschbaum}, {Ottensamer}, {Barlow}, {Blommaert},
  {Gomez}, {Groenewegen}, {Lim}, {Swinyard}, {Waelkens}, \&
  {Tielens}}]{decin12}
{Decin}, L., {Cox}, N.~L.~J., {Royer}, P., {et~al.} 2012, \aap, 548, A113

\bibitem[{Festou(2005)}]{festou05}
Festou, M.~C. 2005, in Comets II, ed. M.~C. Festou, H.~U. Keller, \& H.~A.
  Weaver (Univ. Arizona Press), 3--16

\bibitem[{{Fonfr\'ia} {et~al.}(2022){Fonfr\'ia}, {DeWitt}, {Montiel},
  {Cernicharo}, \& {Richter}}]{fonfria22}
{Fonfr\'ia}, J.~P., {DeWitt}, C.~N., {Montiel}, E.~J., {Cernicharo}, J., \&
  {Richter}, M.~J. 2022, ApJ, submitted, 0

\bibitem[{{Fonfr{\'\i}a} {et~al.}(2021){Fonfr{\'\i}a}, {Montiel}, {Cernicharo},
  {DeWitt}, {Richter}, {Lacy}, {Greathouse}, {Santander-Garc{\'\i}a},
  {Ag{\'u}ndez}, \& {Massalkhi}}]{fonfria21}
{Fonfr{\'\i}a}, J.~P., {Montiel}, E.~J., {Cernicharo}, J., {et~al.} 2021, \aap,
  651, A8

\bibitem[{{Fong} {et~al.}(2006){Fong}, {Meixner}, {Sutton}, {Zalucha}, \&
  {Welch}}]{fong06}
{Fong}, D., {Meixner}, M., {Sutton}, E.~C., {Zalucha}, A., \& {Welch}, W.~J.
  2006, \apj, 652, 1626

\bibitem[{{Glassgold}(1996)}]{glassgold96}
{Glassgold}, A.~E. 1996, \araa, 34, 241

\bibitem[{{Goldreich} \& {Scoville}(1976)}]{goldreich76}
{Goldreich}, P., \& {Scoville}, N. 1976, \apj, 205, 144

\bibitem[{{Goldsmith} {et~al.}(2012){Goldsmith}, {Langer}, {Pineda}, \&
  {Velusamy}}]{goldsmith12}
{Goldsmith}, P.~F., {Langer}, W.~D., {Pineda}, J.~L., \& {Velusamy}, T. 2012,
  \apjs, 203, 13

\bibitem[{{Gorti} \& {Hollenbach}(2004)}]{gorti04}
{Gorti}, U., \& {Hollenbach}, D. 2004, \apj, 613, 424

\bibitem[{{Groenewegen} {et~al.}(2012){Groenewegen}, {Barlow}, {Blommaert},
  {Cernicharo}, {Decin}, {Gomez}, {Hargrave}, {Kerschbaum}, {Ladjal}, {Lim},
  {Matsuura}, {Olofsson}, {Sibthorpe}, {Swinyard}, {Ueta}, \&
  {Yates}}]{groenewegen12}
{Groenewegen}, M.~A.~T., {Barlow}, M.~J., {Blommaert}, J.~A.~D.~L., {et~al.}
  2012, \aap, 543, L8

\bibitem[{{Guan} {et~al.}(2012){Guan}, {Stutzki}, {Graf}, {G{\"u}sten},
  {Okada}, {Requena-Torres}, {Simon}, \& {Wiesemeyer}}]{guan12}
{Guan}, X., {Stutzki}, J., {Graf}, U.~U., {et~al.} 2012, \aap, 542, L4

\bibitem[{{Gu{\'e}lin} {et~al.}(2018){Gu{\'e}lin}, {Patel}, {Bremer},
  {Cernicharo}, {Castro-Carrizo}, {Pety}, {Fonfr{\'\i}a}, {Ag{\'u}ndez},
  {Santander-Garc{\'\i}a}, {Quintana-Lacaci}, {Velilla Prieto}, {Blundell}, \&
  {Thaddeus}}]{guelin18}
{Gu{\'e}lin}, M., {Patel}, N.~A., {Bremer}, M., {et~al.} 2018, \aap, 610, A4

\bibitem[{{H{\"o}fner} \& {Olofsson}(2018)}]{hofner18}
{H{\"o}fner}, S., \& {Olofsson}, H. 2018, \aapr, 26, 1

\bibitem[{{Huggins} {et~al.}(1994){Huggins}, {Bachiller}, {Cox}, \&
  {Forveille}}]{huggins94}
{Huggins}, P.~J., {Bachiller}, R., {Cox}, P., \& {Forveille}, T. 1994, \apjl,
  424, L127

\bibitem[{Hunter(2007)}]{matplotlib}
Hunter, J.~D. 2007, Computing in Science \& Engineering, 9, 90

\bibitem[{{Karakas}(2010)}]{karakas10}
{Karakas}, A.~I. 2010, \mnras, 403, 1413

\bibitem[{{Keene} {et~al.}(1993){Keene}, {Young}, {Phillips}, {Buettgenbach},
  \& {Carlstrom}}]{keene93}
{Keene}, J., {Young}, K., {Phillips}, T.~G., {Buettgenbach}, T.~H., \&
  {Carlstrom}, J.~E. 1993, \apjl, 415, L131

\bibitem[{{Kwan} \& {Linke}(1982)}]{kwan82}
{Kwan}, J., \& {Linke}, R.~A. 1982, \apj, 254, 587

\bibitem[{{Ladjal} {et~al.}(2010){Ladjal}, {Barlow}, {Groenewegen}, {Ueta},
  {Blommaert}, {Cohen}, {Decin}, {De Meester}, {Exter}, {Gear}, {Gomez},
  {Hargrave}, {Huygen}, {Ivison}, {Jean}, {Kerschbaum}, {Leeks}, {Lim},
  {Olofsson}, {Polehampton}, {Posch}, {Regibo}, {Royer}, {Sibthorpe},
  {Swinyard}, {Vandenbussche}, {Waelkens}, \& {Wesson}}]{ladjal10}
{Ladjal}, D., {Barlow}, M.~J., {Groenewegen}, M.~A.~T., {et~al.} 2010, \aap,
  518, L141

\bibitem[{{Li} {et~al.}(2013){Li}, {Heays}, {Visser}, {Ubachs}, {Lewis},
  {Gibson}, \& {van Dishoeck}}]{li13}
{Li}, X., {Heays}, A.~N., {Visser}, R., {et~al.} 2013, \aap, 555, A14

\bibitem[{{Mamon} {et~al.}(1988){Mamon}, {Glassgold}, \& {Huggins}}]{mamon88}
{Mamon}, G.~A., {Glassgold}, A.~E., \& {Huggins}, P.~J. 1988, \apj, 328, 797

\bibitem[{{McWilliam}(1997)}]{mcwilliam97}
{McWilliam}, A. 1997, \araa, 35, 503

\bibitem[{{Menten} {et~al.}(2012){Menten}, {Reid}, {Kami{\'n}ski}, \&
  {Claussen}}]{menten12}
{Menten}, K.~M., {Reid}, M.~J., {Kami{\'n}ski}, T., \& {Claussen}, M.~J. 2012,
  \aap, 543, A73

\bibitem[{{Milam} {et~al.}(2005){Milam}, {Savage}, {Brewster}, {Ziurys}, \&
  {Wyckoff}}]{milam05}
{Milam}, S.~N., {Savage}, C., {Brewster}, M.~A., {Ziurys}, L.~M., \& {Wyckoff},
  S. 2005, \apj, 634, 1126

\bibitem[{{Millar}(2020)}]{millar20}
{Millar}, T.~J. 2020, Chinese Journal of Chemical Physics, 33, 668

\bibitem[{{Millar} {et~al.}(2000){Millar}, {Herbst}, \& {Bettens}}]{millar00}
{Millar}, T.~J., {Herbst}, E., \& {Bettens}, R.~P.~A. 2000, \mnras, 316, 195

\bibitem[{{M{\"u}ller} {et~al.}(2005){M{\"u}ller}, {Schl{\"o}der}, {Stutzki},
  \& {Winnewisser}}]{mueller05}
{M{\"u}ller}, H. S.~P., {Schl{\"o}der}, F., {Stutzki}, J., \& {Winnewisser}, G.
  2005, Journal of Molecular Structure, 742, 215

\bibitem[{{Nieva} \& {Przybilla}(2012)}]{nieva12}
{Nieva}, M.~F., \& {Przybilla}, N. 2012, \aap, 539, A143

\bibitem[{{Olofsson} {et~al.}(2015){Olofsson}, {Bergman}, \&
  {Lindqvist}}]{oloffson15}
{Olofsson}, H., {Bergman}, P., \& {Lindqvist}, M. 2015, \aap, 582, A102

\bibitem[{{Ossenkopf} {et~al.}(2013){Ossenkopf}, {R{\"o}llig}, {Neufeld},
  {Pilleri}, {Lis}, {Fuente}, {van der Tak}, \& {Bergin}}]{ossenkopf13}
{Ossenkopf}, V., {R{\"o}llig}, M., {Neufeld}, D.~A., {et~al.} 2013, \aap, 550,
  A57

\bibitem[{{Pilbratt} {et~al.}(2010){Pilbratt}, {Riedinger}, {Passvogel},
  {Crone}, {Doyle}, {Gageur}, {Heras}, {Jewell}, {Metcalfe}, {Ott}, \&
  {Schmidt}}]{pilbratt10}
{Pilbratt}, G.~L., {Riedinger}, J.~R., {Passvogel}, T., {et~al.} 2010, \aap,
  518, L1

\bibitem[{{Poglitsch} {et~al.}(2010){Poglitsch}, {Waelkens}, {Geis},
  {Feuchtgruber}, {Vandenbussche}, {Rodriguez}, {Krause}, {Renotte}, {van
  Hoof}, {Saraceno}, {Cepa}, {Kerschbaum}, {Agn{\`e}se}, {Ali}, {Altieri},
  {Andreani}, {Augueres}, {Balog}, {Barl}, {Bauer}, {Belbachir}, {Benedettini},
  {Billot}, {Boulade}, {Bischof}, {Blommaert}, {Callut}, {Cara}, {Cerulli},
  {Cesarsky}, {Contursi}, {Creten}, {De Meester}, {Doublier}, {Doumayrou},
  {Duband}, {Exter}, {Genzel}, {Gillis}, {Gr{\"o}zinger}, {Henning},
  {Herreros}, {Huygen}, {Inguscio}, {Jakob}, {Jamar}, {Jean}, {de Jong},
  {Katterloher}, {Kiss}, {Klaas}, {Lemke}, {Lutz}, {Madden}, {Marquet},
  {Martignac}, {Mazy}, {Merken}, {Montfort}, {Morbidelli}, {M{\"u}ller},
  {Nielbock}, {Okumura}, {Orfei}, {Ottensamer}, {Pezzuto}, {Popesso},
  {Putzeys}, {Regibo}, {Reveret}, {Royer}, {Sauvage}, {Schreiber}, {Stegmaier},
  {Schmitt}, {Schubert}, {Sturm}, {Thiel}, {Tofani}, {Vavrek}, {Wetzstein},
  {Wieprecht}, \& {Wiezorrek}}]{poglitsch10}
{Poglitsch}, A., {Waelkens}, C., {Geis}, N., {et~al.} 2010, \aap, 518, L2

\bibitem[{{Ramstedt} {et~al.}(2020){Ramstedt}, {Vlemmings}, {Doan},
  {Danilovich}, {Lindqvist}, {Saberi}, {Olofsson}, {De Beck}, {Groenewegen},
  {H{\"o}fner}, {Kastner}, {Kerschbaum}, {Khouri}, {Maercker}, {Montez},
  {Quintana-Lacaci}, {Sahai}, {Tafoya}, \& {Zijlstra}}]{ramstedt20}
{Ramstedt}, S., {Vlemmings}, W.~H.~T., {Doan}, L., {et~al.} 2020, \aap, 640,
  A133

\bibitem[{{Risacher} {et~al.}(2016){Risacher}, {G{\"u}sten}, {Stutzki},
  {H{\"u}bers}, {Bell}, {Buchbender}, {B{\"u}chel}, {Csengeri}, {Graf},
  {Heyminck}, {Higgins}, {Honingh}, {Jacobs}, {Klein}, {Okada}, {Parikka},
  {P{\"u}tz}, {Reyes}, {Ricken}, {Riquelme}, {Simon}, \&
  {Wiesemeyer}}]{risacher16}
{Risacher}, C., {G{\"u}sten}, R., {Stutzki}, J., {et~al.} 2016, \aap, 595, A34

\bibitem[{{Risacher} {et~al.}(2018){Risacher}, {G{\"u}sten}, {Stutzki},
  {H{\"u}bers}, {Aladro}, {Bell}, {Buchbender}, {B{\"u}chel}, {Csengeri},
  {Duran}, {Graf}, {Higgins}, {Honingh}, {Jacobs}, {Justen}, {Klein},
  {Mertens}, {Okada}, {Parikka}, {P{\"u}tz}, {Reyes}, {Richter}, {Ricken},
  {Riquelme}, {Rothbart}, {Schneider}, {Simon}, {Wienold}, {Wiesemeyer},
  {Ziebart}, {Fusco}, {Rosner}, \& {Wohler}}]{risacher18}
---. 2018, Journal of Astronomical Instrumentation, 7, 1840014

\bibitem[{{Saberi} {et~al.}(2019){Saberi}, {Vlemmings}, \& {De
  Beck}}]{saberi19}
{Saberi}, M., {Vlemmings}, W.~H.~T., \& {De Beck}, E. 2019, \aap, 625, A81

\bibitem[{{Saberi} {et~al.}(2018){Saberi}, {Vlemmings}, {De Beck}, {Montez}, \&
  {Ramstedt}}]{saberi18}
{Saberi}, M., {Vlemmings}, W.~H.~T., {De Beck}, E., {Montez}, R., \&
  {Ramstedt}, S. 2018, \aap, 612, L11

\bibitem[{{Sahai} \& {Chronopoulos}(2010)}]{sahai10}
{Sahai}, R., \& {Chronopoulos}, C.~K. 2010, \apjl, 711, L53

\bibitem[{{Savage} {et~al.}(1977){Savage}, {Bohlin}, {Drake}, \&
  {Budich}}]{savage77}
{Savage}, B.~D., {Bohlin}, R.~C., {Drake}, J.~F., \& {Budich}, W. 1977, \apj,
  216, 291

\bibitem[{{Shipman} {et~al.}(2017){Shipman}, {Beaulieu}, {Teyssier}, {Morris},
  {Rengel}, {McCoey}, {Edwards}, {Kester}, {Lorenzani}, {Coeur-Joly},
  {Melchior}, {Xie}, {Sanchez}, {Zaal}, {Avruch}, {Borys}, {Braine}, {Comito},
  {Delforge}, {Herpin}, {Hoac}, {Kwon}, {Lord}, {Marston}, {Mueller}, {Olberg},
  {Ossenkopf}, {Puga}, \& {Akyilmaz-Yabaci}}]{shipman17}
{Shipman}, R.~F., {Beaulieu}, S.~F., {Teyssier}, D., {et~al.} 2017, \aap, 608,
  A49

\bibitem[{{Smith} {et~al.}(2014){Smith}, {Glover}, {Clark}, {Klessen}, \&
  {Springel}}]{smith14}
{Smith}, R.~J., {Glover}, S.~C.~O., {Clark}, P.~C., {Klessen}, R.~S., \&
  {Springel}, V. 2014, \mnras, 441, 1628

\bibitem[{{Temi} {et~al.}(2018){Temi}, {Hoffman}, {Ennico}, \& {Le}}]{temi18}
{Temi}, P., {Hoffman}, D., {Ennico}, K., \& {Le}, J. 2018, Journal of
  Astronomical Instrumentation, 7, 1840011

\bibitem[{{Teyssier} \& {Rengel}(2016)}]{teyssier16}
{Teyssier}, D., \& {Rengel}, M. 2016, HIFI Reference Position Spectra Data
  Products: Release notes, herschel-hsc-doc-2111 edn., European Space Agency

\bibitem[{{Tielens} \& {Hollenbach}(1985)}]{tielens85}
{Tielens}, A.~G.~G.~M., \& {Hollenbach}, D. 1985, \apj, 291, 722

\bibitem[{{Ueta} {et~al.}(2019){Ueta}, {Torres}, {Izumiura}, {Yamamura},
  {Takita}, \& {Tomasino}}]{ueta19}
{Ueta}, T., {Torres}, A.~J., {Izumiura}, H., {et~al.} 2019, \pasj, 71, 4

\bibitem[{{Visser} {et~al.}(2009){Visser}, {van Dishoeck}, \&
  {Black}}]{visser09}
{Visser}, R., {van Dishoeck}, E.~F., \& {Black}, J.~H. 2009, \aap, 503, 323

\bibitem[{{Wakelam} {et~al.}(2015){Wakelam}, {Loison}, {Herbst}, {Pavone},
  {Bergeat}, {B{\'e}roff}, {Chabot}, {Faure}, {Galli}, {Geppert}, {Gerlich},
  {Gratier}, {Harada}, {Hickson}, {Honvault}, {Klippenstein}, {Le Picard},
  {Nyman}, {Ruaud}, {Schlemmer}, {Sims}, {Talbi}, {Tennyson}, \&
  {Wester}}]{wakelam15}
{Wakelam}, V., {Loison}, J.~C., {Herbst}, E., {et~al.} 2015, \apjs, 217, 20

\bibitem[{{Willson}(2000)}]{wilson00}
{Willson}, L.~A. 2000, \araa, 38, 573

\bibitem[{{Wood} {et~al.}(2019){Wood}, {Smythe}, \& {Harrison}}]{wood19}
{Wood}, B.~J., {Smythe}, D.~J., \& {Harrison}, T. 2019, Amer. Mineralogist,
  104, 844

\bibitem[{{Young} {et~al.}(2012){Young}, {Becklin}, {Marcum}, {Roellig}, {De
  Buizer}, {Herter}, {G{\"u}sten}, {Dunham}, {Temi}, {Andersson}, {Backman},
  {Burgdorf}, {Caroff}, {Casey}, {Davidson}, {Erickson}, {Gehrz}, {Harper},
  {Harvey}, {Helton}, {Horner}, {Howard}, {Klein}, {Krabbe}, {McLean}, {Meyer},
  {Miles}, {Morris}, {Reach}, {Rho}, {Richter}, {Roeser}, {Sandell}, {Sankrit},
  {Savage}, {Smith}, {Shuping}, {Vacca}, {Vaillancourt}, {Wolf}, \&
  {Zinnecker}}]{young12}
{Young}, E.~T., {Becklin}, E.~E., {Marcum}, P.~M., {et~al.} 2012, \apjl, 749,
  L17

\end{thebibliography}

\end{document}